\documentclass[final,1p,times,twocolumn]{elsarticle}

\usepackage{epsfig}

\usepackage{amssymb}
\usepackage{amsthm}
\usepackage{amsmath}
\usepackage{amsbsy}
\usepackage{widetext}
\usepackage{morefloats}

\begin{document}
\begin{frontmatter}



\title{Concept and Analysis of a Satellite for Space-based Radio Detection of Ultra-high Energy Cosmic Rays}


\author[label1]{Andrew Romero-Wolf}
\author[label2]{Peter Gorham}
\author[label1]{Kurt Liewer}
\author[label1]{Jeffrey Booth}
\author[label1]{Riley Duren}

\address[label1]{Jet Propulsion Laboratory, California Institute of Technology, 4800 Oak Grove Drive, Pasadena, California 91109, USA}
\address[label2]{Department of Physics and Astronomy, University of Hawai'i at Manoa, Honolulu, HI 96822, USA}


\begin{abstract}
We present a concept for on-orbit radio detection of ultra-high energy cosmic rays (UHECRs) that has the potential to provide collection rates of $\sim$100 events per year for energies above $10^{20}$~eV. The synoptic wideband orbiting radio detector (SWORD) mission's high event statistics at these energies combined with the pointing capabilities of a space-borne antenna array could enable charged particle astronomy. The detector concept is based on ANITA's successful detection UHECRs where the geosynchrotron radio signal produced by the extended air shower is reflected off the Earth's surface and detected in flight. 
\end{abstract}


\end{frontmatter}



\section{Introduction}
The origin of ultra-high energy cosmic rays (UHECRs) remains a mystery decades after their discovery~\cite{Auger_2010, Kotera_2011, Letessier_2011}. Deflection due to cosmic magnetic fields impedes the reconstruction of source direction except for the highest energies where current observations are statistically limited. A key to identifying the sources of UHECRs is to increase the exposure at energies above $6\times10^{19}$~eV. We present a space-based mission concept using radio detection techniques with the potential to increase the exposure by factors of 40-70 over Auger observatory at super-GZK energies relevant to charged particle astronomy.

Radio detection efforts for UHECRs date to the 1960's~\cite{jelley_1965} and remained active through the 1970's~\cite{cr_1,cr_2,cr_3,cr_4,cr_5,cr_6,cr_7,cr_8,cr_9}. The technique originally promised an alternative to particle counters on the ground with a wider field of view. However, the effort was abandoned due to technological limitations. The field has had a relatively recent rebirth~\cite{falcke_2005} with the advent of low power fast digitizers. Existing ground arrays are currently adding improved radio detection capabilities \cite{nehls_2008, abreu_2012}. 

In a different radio detection effort, the ANITA experiment was developed as a balloon-borne antenna array designed to detect radio pulses in the 200-1200 MHz frequency range from ultra-high energy neutrinos interacting in the Antarctic ice~\cite{gorham_2009b}. No neutrinos were detected but there was a serendipitous observation of UHECR pulses reflected off the ice-sheet~\cite{hoover_2010}. 

\begin{figure*}[t]
\centering
\includegraphics[width=0.7\linewidth]{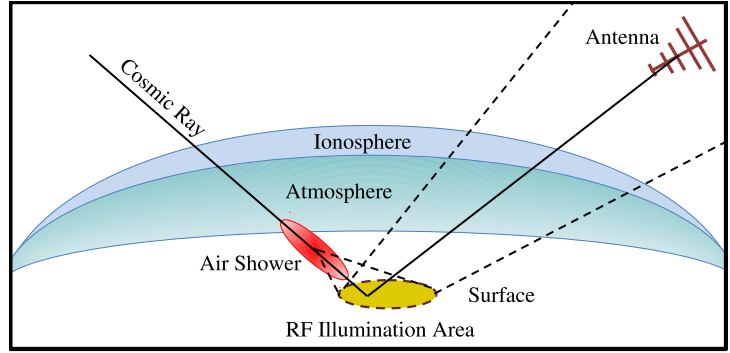}
\caption{Diagram of the SWORD detection concept. The cosmic ray comes in from outer space and interacts with the atmosphere to produce an extended air shower. The geomagnetic field induces a charge separation to produce a forward directed geosynchrotron pulse which illuminates a spot on the surface of the Earth. The reflected pulse propagates though the atmosphere and the ionosphere to be detected by an antenna in low Earth orbit.} 
\label{fig:concept} 
\end{figure*}

The Synoptic Wideband Orbiting Radio Detector (SWORD) is a mission based on the ANITA detection of UHECRs~\cite{hoover_2010}. SWORD detects the geosynchrotron radiation~\cite{falcke_2003, suprun_2003} of UHECR extended air showers (EAS) produced in the atmosphere, reflected off the Earth's surface, and detected from low Earth orbit (LEO) (see Figure~\ref{fig:concept}). Orbit altitudes (600-800~km) provide a large collection area for UHECRs with energies $>10^{20}$~eV with detection rates of $\sim$100 per year. However, these altitudes also require that the cosmic ray induced radio pulse propagate through the ionosphere, which is a major technological challenge for the success of this mission and is addressed in detail in this study.

In this paper we describe the concept and analysis of SWORD. Section 2 presents the signal simulations based on a geosynchrotron radiation parameterization including models for oblique reflections, surface roughness, noise backgrounds, and dispersion the radio impulse through the ionosphere. Section 3 describes the mission architecture which entails the choice of antenna and triggering of pulses dispersed through the ionosphere. Section 4 gives the expected performance of the mission based on detailed simulations. In Section 5 we give the conclusions of our results and the outlook for developing this mission.

\section{Signal Simulation}
In this section we provide a parametrized model of the signal production from geosynchrotron radiation with propagation effects due to reflection on the surface of the Earth, including surface roughness, and the ionosphere. The noise contributions of the sky, ground, and anthropogenic backgrounds are estimated. 
\subsection{Geosynchrotron Radiation}
Geosynchrotron radiation is produced by the electrons and positrons in the extended air shower interacting with the geomagnetic field. The Lorentz force separates the charges causing an overall transverse current~\cite{falcke_2003,suprun_2003}. The total number of electrons and positrons in the shower propagates in a meter-scale longitudinal thickness `pancake' resulting in a radio pulse that is coherent at frequencies below 100 MHz with partial coherence up to $\lesssim$1~GHz~\cite{hoover_2010}. At these frequencies, the radio impulse power is proportional to the square of the shower energy resulting in strongly detectable pulses over large distances at the highest energies. 

The parametrization of geosynchrotron radiation applied in this study is based on the results of a data-driven maximum-likelihood model applied to the ANITA UHECR events~\cite{hoover_2010}. Despite steady progress in radio emission calculations over the last decade~\cite{huege_2005, ludwig_2011, huege_2013, scholten_2008, alvarez_2012a, alvarez_2012b} these models do not directly apply to the geometry of a detector in flight where observations are in the far-field, in contrast to ground arrays. Observations from the sky also involve much larger zenith angles than ground arrays usually observe. The parametrization applied to the ANITA data and for the SWORD simulations is given by

\begin{equation}
\mathbf{\mathcal{E}}_{p}(f)= A_{0}
\left(\frac{E_{sh}}{10^{19}\mbox{ eV}}\right)
\left(\frac{R_{ref}}{R+R_{Xmax}}\right)
F(f,\theta_{sh})
S\left(f\right)
\frac{\mathbf{B}_{\perp}\cdot\mathbf{\hat{p}}}{B_{ref}}
\cos(\theta_z)
\mathcal{F}_p(\theta_z)
G(f,R)
\label{eq:geo_synch_param}
\end{equation}
where

\begin{tabular}{ll}
$\mathbf{\mathcal{E}}_{p}(f)$ & is the Fourier component of the electric field vector,\\
$f$ & is the frequency,\\
$p$ & is the linear polarization vector component with unit vector $\mathbf{\hat{p}}$,\\
$A_{0}$  & is the reference amplitude given in $\mu$V/m/MHz, \\
$E_{sh}$ & is the shower energy, \\
$R_{ref}$ & is a reference distance on the ground, \\
$R_{Xmax}$ & is the distance of shower maximum to the ground, \\
$R$ & is the distance of the observer to the ground, \\
$\theta_{sh}$ & is the angle of observation with respect to the shower axis, \\
$F(f,\theta_{sh})$ & is the radio beam of the extended air shower with its peak normalized to unity. \\
$S(f)$ & is equal to $\exp[(265\mbox{ MHz}-f)/ 365\mbox{ MHz}]$ for f$>$100 MHz and $\exp(-165/365)$ for f$<$100 MHz, \\
$\mathbf{B}_{\perp}$ & is the cross product of the the geomagnetic field with the shower axis, \\
$B_{ref}$ & is the reference value of the geomagnetic field, \\
$\theta_z$ & is the zenith angle of incidence of the shower on the ground, \\
$\cos(\theta_z)$ & is the obliquity factor due to reflection, \\
$\mathcal{F}_p(\theta_z)$ & is the Fresnel reflection coefficient for polarization component $p$, and \\
$G(f,R)$ & is the surface roughness factor. \\
\end{tabular}
\linebreak

The reference amplitude used in this parameterization is determined from the ANITA data at $360^{+100}_{-250}$ $\mu$V/m/MHz~\cite{hoover_2010}. The values of $A_{0}$ for parameterizations similar to that presented here vary widely in the literature. For our energy scaling with respect to $10^{19}$~eV, corrected for the exponentially falling spectrum, $A_0$ ranges from as low as 30~$\mu$V/m/MHz~\cite{prah_1971} to as high as 390~$\mu$V/m/MHz~\cite{huege_2008} with values in between lying closer to the smaller value~\cite{cr_9, huege_2005, lopes_2009}.

The $R_{ref}/(R+R_{Xmax})$ term accounts for the $1/R$ propagation loss of a radiative electric field including the distance $R_{Xmax}$ from shower maximum to the ground and the distance $R$ from the ground to the detector. The reference distance $R_{ref}=8$~km corresponds to the distance from shower maximum to the ground for a shower with $\theta_z=60^\circ$.

The beam width of the UHECR emission is based on the synchrotron radiation formula found in~\cite{jackson_1999} with parameters fitted to the ANITA data~\cite{hoover_2010}. The beam pattern parametrization $F(f,\theta_{sh})$ depends on the frequency $f$, the angle of observation with respect to shower axis $\theta_{sh}$, and is normalized so that its peak value is unity. The formula given in~\cite{jackson_1999} for the intensity frequency spectrum for a differential solid angle $d\Omega$ is
\begin{equation}
\frac{d^2I}{d\omega d\Omega}\approx\frac{e^2}{3\pi^2 c}\left(\frac{\omega\rho}{c}\right)^2\left(\frac{1}{\gamma^2}+\theta^2\right)^2 
\left[K^2_{2/3}(\xi) + \frac{\theta^2}{1/\gamma^2+\theta^2}K^2_{1/3}(\xi)\right]
\label{eqn:f1}
\end{equation}
where 
\begin{equation}
\xi=\left(\frac{\omega\rho}{3c}\right)\left(\frac{1}{\gamma^2}+\theta^2\right)^{3/2},
\end{equation}
$c$ is the speed of light in vacuum, $\gamma$ is the Lorentz factor, $\theta$ is the angle of observation (in this case $\theta_{sh}$), $\rho$ is the radius of curvature of the particle, and $K_{2/3}(\xi)$ and $K_{1/3}(\xi)$ are modified Bessel functions. The beam parameterization is normalized to the spectrum at $\theta=0$ and given by
\begin{equation}
F(f,\theta) = \left(\sqrt{\left(\left.\frac{d^2I}{d\omega d\Omega}\right / \left.\frac{d^2I}{d\omega d\Omega}\right|_{\theta=0} \right)}\right)^n,
\label{eqn:f2}
\end{equation}
where $n$ is an exponent that is fit to match the ANITA data. Inserting Equation~\ref{eqn:f1} into Equation~\ref{eqn:f2} reduces $F$ to
\begin{equation}
F(f,\theta) = \left(\left(1+\gamma^2\theta^2\right)\sqrt{\frac{K^2_{2/3}(\xi) + \frac{\theta^2}{1/\gamma^2+\theta^2}K^2_{1/3}(\xi)}{K^2_{2/3}(\xi_0) + \gamma^2 K^2_{1/3}(\xi_0)}}\right)^n
\end{equation}
where $\xi_0=\omega\rho/(3c\gamma^2)$

\begin{figure*}[ht!]
\centering
\includegraphics[width=0.7\linewidth]{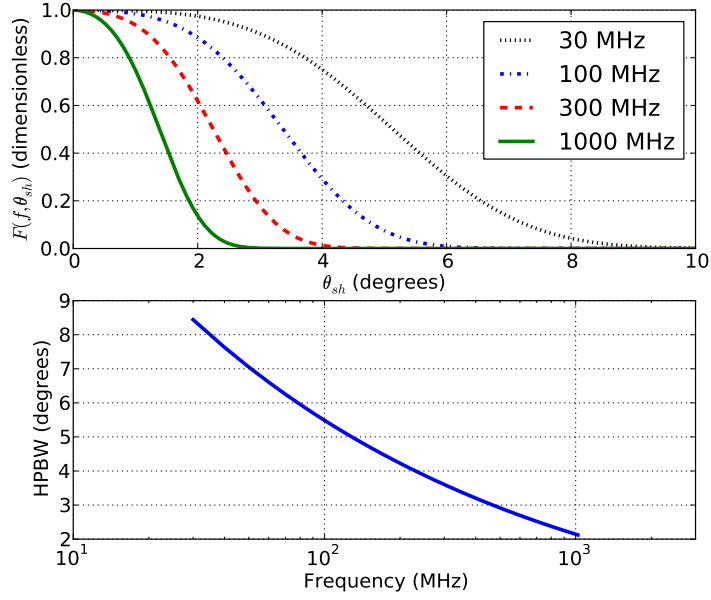}
\caption{The geosynchrotron beam pattern for various frequencies. Shown on top are the $F(f,\theta_{sh})$ factors in Equation~\ref{eq:geo_synch_param}, which gives the beam pattern as a function of angle with respect to the shower axis. The function is normalized to the peak emission. The plot on bottom shows the half-power beam width (HPBW) as a function of frequency. } 
\label{fig:beam} 
\end{figure*}

The radius of curvature for an electron or position under the influence of the geomagnetic field $B\sim 50$~$\mu T$ with $\gamma\sim60$ gives $\rho\approx mc\gamma\beta/eB\approx2$~km (assuming a $\gamma$=60)~\cite{hoover_2010}. At frequencies between 30~MHz and 1~GHz and at observation angles of order $\theta\sim1^{\circ}$, $\xi$ ranges between $6\times10^{-4}$ and $10^{-5}$. If $\xi \ll 1$ then $K_{1/3}(\xi) \ll K_{2/3}(\xi)$ giving
\begin{equation}
F(f,\theta) = \left(\left(1+\gamma^2\theta^2\right)\frac{K_{2/3}(\xi)}{K_{2/3}(\xi_0)}\right)^n.
\end{equation}
With the values of $\gamma\sim60$ and $B\sim50$~$\mu$T, we have the parameterization
\begin{equation}
F(f,\theta_{sh}) = 
\left(
\left(1+1.1\left(\frac{\theta_{sh}}{1^{\circ}}\right)^2\right) 
\frac{K_{2/3}\left[6.6\times 10^{-5} \left(\frac{f}{\mbox{\small MHz}}\right) \left(1+1.1\left(\frac{\theta_{sh}}{1^{\circ}}\right)^2\right)^{3/2}\right]}  
{K_{2/3}\left[6.6\times 10^{-5} \left(\frac{f}{\mbox{\small MHz}}\right) \right]}
\right)^{3.333},
\end{equation}
where $n=3.333$ is the exponent that best fits the ANITA data. Figure~\ref{fig:beam} shows the beam patterns as a function of $\theta_{sh}$ for selected frequencies as well as the half power beam width of the parameterization as a function of frequency over the range 30-300~MHz.

The spectral shape $S(f)$ follows an exponential drop according to the ANITA observations in the range of 300-1000~MHz~\cite{hoover_2010}. The extension to lower frequencies is guided by modern simulations~\cite{alvarez_2012a, alvarez_2012b} that predict that the spectrum of inclined showers, which SWORD is most sensitive to, flattens out below 100~MHz. It is unclear whether this feature remains for far-field observations such as ANITA or SWORD but the spectral flattening assumed in the simulations is a conservative approach and the results would only improve if the signal strength continues to rise with decreasing frequency. 

The $\mathbf{B}_{\perp}$ term is proportional to the cross product of the shower axis with the geomagnetic field. The shower strength varies with the magnetic field intensity, which is referenced to $B_{ref}=45$~$\mu T$.

\subsection{Oblique Reflections}
The remaining terms of the parameterization in Equation~\ref{eq:geo_synch_param} account for reflection effects on the surface. The reflection is calculated using the scalar Kirchoff theory of diffraction. With this formalism we estimate the frequency and distance dependent coefficients for oblique reflection of radio signals off the surface of the Earth. The relevance of this approach will become evident in the treatment of surface roughness.

\begin{figure*}[h!]
\centering
\includegraphics[width=0.7\linewidth]{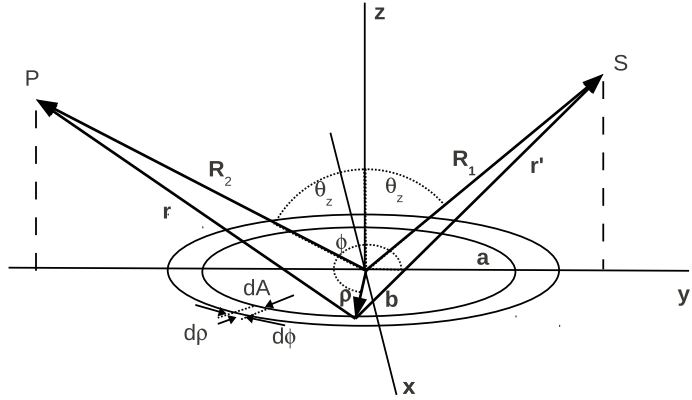}
\caption{ Geometry for the reflections calculated using Equation~\ref{eqn:Kirchoff_on_ellipse}. This is an adaptation of the standard Kirchoff-Fresnel diffraction through and opening in an opaque plane. The illumination from source $S$ has a projected ellipse on the ground which is reflected to a receiver at point $P$. The Kirchoff-Fresnel integral includes contributions not just from the specular reflection path, given by vectors $\mathbf{R}_1$ and $\mathbf{R}_2$, but from all paths on the projected area given by vectors $\mathbf{r}$ and $\mathbf{r}'$. See text for details.} 
\label{fig:reflection_geometry} 
\end{figure*}

The geometry of the oblique reflection is shown in Figure~\ref{fig:reflection_geometry}. The source is located at point $S$ and the receiver at point $P$. The origin is located at the specular reflection point and the position vectors of the source and payload are $\mathbf{R}_1$ and $\mathbf{R}_2$, respectively. The zenith angles for each position vector are $\theta_z$ given by the specular reflection condition in the plane of incidence. The position vector {\boldmath$\rho$}=($\rho$,$\phi$) of an area element on the surface is $dA$. The vector from this element to the source is $\mathbf{r}'=\mathbf{R}_1-\mbox{\boldmath$\rho$}$ and to the receiver is $\mathbf{r}=\mathbf{R}-\mbox{\boldmath$\rho$}$. In the Kirchoff theory of diffraction, the signal at point $P$ has contributions from reflections of the source at $S$ over the whole surface.

The region over which the path length difference is less than some fixed value will be bounded by an ellipse on a flat surface, with major radius and minor radius b, satisfying the condition $a=b/\cos\theta_z$. Figure~\ref{fig:reflection_geometry}, shows an annular region with thickness $d\rho$. The problem as formulated so far is the same as determining the size of that region in a mirror around the specular point which contributes to the far-field reflection. In our case, however, the reflection coefficient is not that of a perfect conductor, so we must use the Fresnel coefficients for reflection from a dielectric surface.

For this analysis we assume that $\rho\ll R_1,R_2$ so that the polarization vector is approximately constant across the reflective region, and we can use a constant Fresnel reflection coefficient for each component of the polarization. Using the Kirchoff approximation~\cite{jackson_1999}, the received field strength $E_{rcv}$ for monochromatic waves at $P$ due to a source field strength $E_{src}$ at point $S$ is given by
\begin{equation}
E^p_{rcv}=\frac{k}{2\pi i}\int_{surface}E^p_{src}(\omega,\theta)\mathcal{F}_p(\theta)\frac{e^{ik|\mathbf{r}|}}{|\mathbf{r}|}\frac{e^{ik|\mathbf{r}'|}}{|\mathbf{r}'|}\cos\theta dA
\label{eqn:kirchoff}
\end{equation}
where $k=2\pi/\lambda$ is the wavenumber, the index $p$ refers to the polarization vector component in the transverse electric or transverse magnetic polarization, and the corresponding Fresnel coefficient $\mathcal{F}_p(\theta)$ for a given zenith angle $\cos\theta=\mathbf{\hat{z}\cdot r}'$ at $dA$. Let us ignore the effects of a curved Earth for now and assume that the reflection surface is planar. 

As shown in Figure~\ref{fig:reflection_geometry} we evaluate the integral over the projected ellipse of the source since this is the shape the UHECR radio illumination takes on the ground. For an ellipse with semi-minor radius $b$, given by $x^2+y^2\cos^2\theta_z=b^2$, Equation~\ref{eqn:kirchoff} becomes
\begin{equation}
E^p_{rcv}=
\frac{k}{2\pi i}
\int_{-b/\cos\theta_z}^{-b/\cos\theta_z}dy
\int_{-\sqrt{b^2-y^2\cos^2\theta_z}}^{\sqrt{b^2-y^2\cos^2\theta_z}} dx
E^p_{src}(\omega,\theta)\mathcal{F}_p(\theta)\frac{e^{ik|\mathbf{r}|}}{|\mathbf{r}|}\frac{e^{ik|\mathbf{r}'|}}{|\mathbf{r}'|}\cos\theta
\label{eqn:Kirchoff_on_ellipse}
\end{equation}

Let us assume that $b$ is small enough so that $\theta\approx\theta_z$ over the region of integration and the half-power beam-width (HPBW) of the UHECR beam is approximately constant. The distances $r\approx R_1$ and $r'\approx R_2$ in the denominator of the integrand but not in the argument of the phasors. Equation \ref{eqn:Kirchoff_on_ellipse} becomes
\begin{equation}
E^p_{rcv}=
E^p_{src}(\omega)
\mathcal{F}_p(\theta_z)
\frac{k}{2\pi i}
\frac{\cos\theta_z}{R_1 R_2}
\int_{-b/\cos\theta_z}^{-b/\cos\theta_z}dy
\int_{-\sqrt{b^2-y^2\cos^2\theta_z}}^{\sqrt{b^2-y^2\cos^2\theta_z}} dx
e^{ik|\mathbf{r}|}e^{ik|\mathbf{r}'|}.
\label{eqn:area_integral}
\end{equation}

\begin{figure*}[ht!]
\centering
\includegraphics[width=0.9\linewidth]{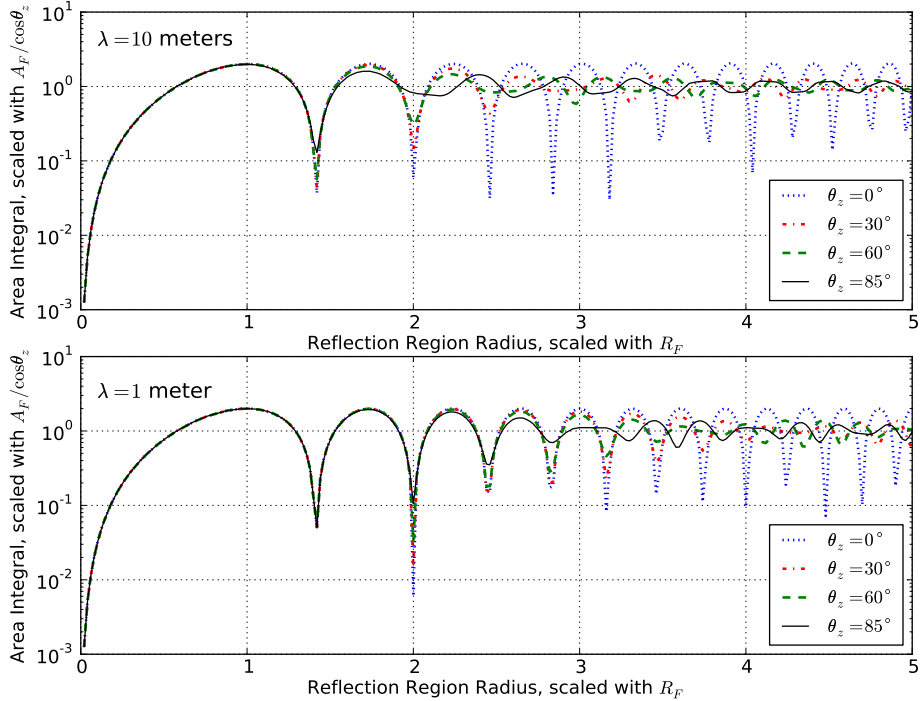}
\caption{ The area integral of Equation~\ref{eqn:area_integral} evaluated for various zenith angles. Equation~\ref{eqn:area_integral} is an approximation of the Kirchoff-Fresnel surface integral of Equation~\ref{eqn:kirchoff} where the integrand consists of the phasors corresponding to the different path lengths light can travel. The result shows that only the first Fresnel zone, projected on the ground, contributes to the magnitude of the integral. The oscillations due to diffraction become more stable over a wider region of integration. } 
\label{fig:area_integral} 
\end{figure*}

\begin{figure*}[h!]
\centering
\includegraphics[width=0.7\linewidth]{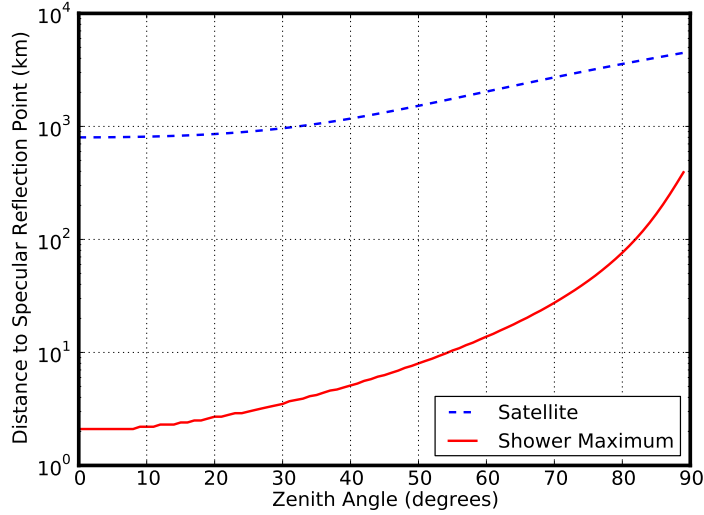}
\caption{ Distance to the specular reflection point as a function of zenith angle for the cosmic ray shower maximum and for a satellite at 800~km orbit altitude.} 
\label{fig:specular_distance} 
\end{figure*}

In this situation, the calculation reduces to an area integral weighted by the product of the two phasors $e^{ik|\mathbf{r}|}$ and  $e^{ik|\mathbf{r'}|}$. In the case of propagation of a free spherical wave, the equivalent integral gives approximately the area of the Fresnel patch~\cite{hecht_1987}
\begin{equation}
A_F = \frac{\lambda R_2 R_1}{R_1+R_2}.
\end{equation}
We can estimate the degree to which the area integral approximates $A_F$ by evaluating the integral in Equation~\ref{eqn:area_integral} numerically. Figure~\ref{fig:area_integral} shows the numerical evaluation of the integral in Equation~\ref{eqn:area_integral}, scaled by the area of the first Fresnel zone projected on the ground $A_F/\cos\theta_z$, evaluated at the Fresnel distance $R_F=\sqrt{\lambda R_1 R_2 / (R_1 + R_2)}$. The $1/\cos\theta_z$ factor is due to the oblique projection to the ellipse. For high zenith angles, the integral converges to $A_F/\cos\theta_z$ after the first couple of Fresnel distances $R_F$. The integral oscillates more as the zenith angle is decreased to vertical where the source distance to the reflection point is smaller (see Figure~\ref{fig:area_integral}). The distance from the specular reflection point to the UHECR shower maximum, assuming an average shower maximum depth of 725~g/cm$^2$, and to the satellite, assuming an altitude of 800~km, are shown in Figure~\ref{fig:specular_distance}. For zenith angles above 30$^{\circ}$, which are of interest to SWORD, it is safe to assume that the magnitude of the integral is given by $A_F/\cos\theta_z$. The magnitude of Equation~\ref{eqn:area_integral} then becomes
\begin{equation}
E^p_{rcv}\approx
E^p_{src}(\omega)
\frac{\mathcal{F}_p(\theta_z)}{R_1+R_2},
\end{equation}
as expected from far-field specular reflection. The oscillations shown in Figure~\ref{fig:area_integral} are of the order of 10\% after a few Fresnel distances. These oscillations are less pronounced for high bandwidth signals such as UHECR pulses compared to the monochromatic estimates shown. 

\begin{figure*}[h!]
\centering
\includegraphics[width=0.7\linewidth]{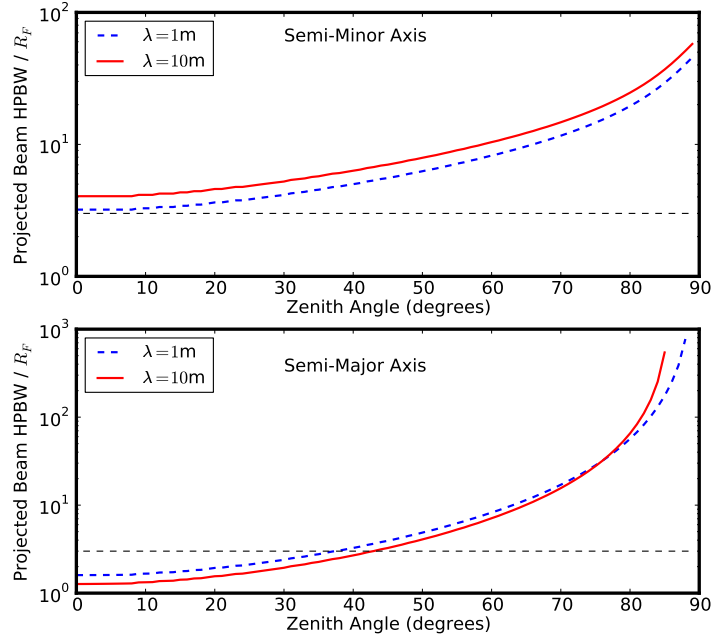}
\caption{ Ratio of the projected UHECR beam HPBW (see Figure~\ref{fig:beam}) ellipse semi-major and semi-minor axes to the Fresnel distance $R_F=\sqrt{\lambda R_1 R_2/(R_1+R_2)}$ calculated from the distances to the specular reflection point from shower maximum and from a receiver in orbit at 800~km altitude (see Figure~\ref{fig:reflection_geometry}). The horizontal dashed line marks a ratio of three, above which value the Kirchoff-Fresnel integral is stable (see Figure~\ref{fig:area_integral}). } 
\label{fig:projected_beam_ratio} 
\end{figure*}

The last piece to justifying these approximations is to compare the size of the UHECR radio illumination, projected on the surface of the Earth, to the size of the Fresnel zone. Figure~\ref{fig:projected_beam_ratio} shows that ratio of the projected UHECR beam HPBW ellipse, for both the semi-major and semi-minor axes, to the Fresnel distance $R_F$. The size of the ellipse is in all cases greater than $R_F$. For projected HPBW to $R_F$ ratios greater than 3, where the area integral converges, the assumption that $E_{src}$ is constant in Equation~\ref{eqn:Kirchoff_on_ellipse} is well justified. Greater care may be needed for ratios lower than three but, as will be shown, there are not a lot of events in the SWORD field of view for the low zenith angles where this is the case.

\begin{figure*}[h!]
\centering
\includegraphics[width=0.7\linewidth]{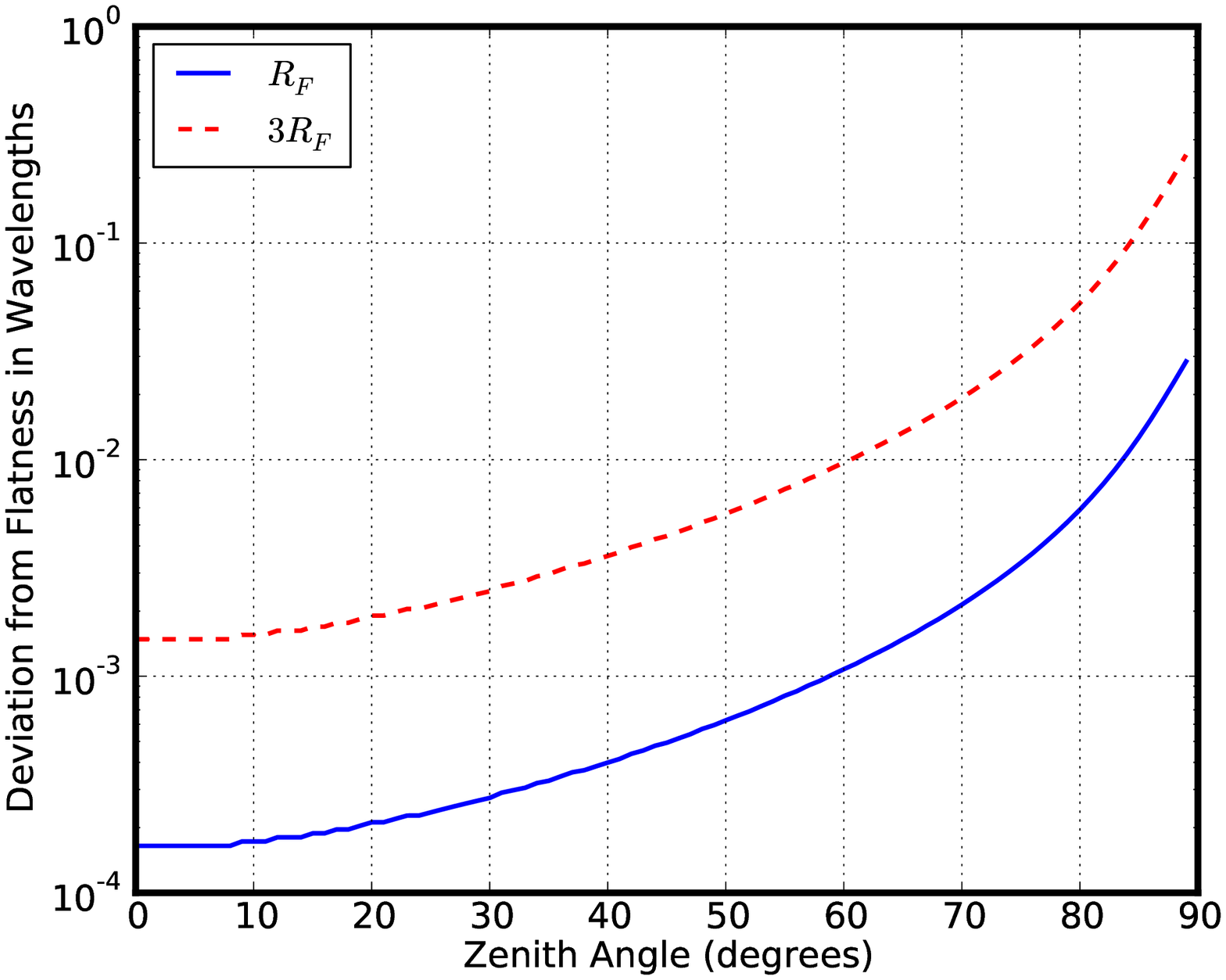}
\caption{ The deviation from flatness for a distance Fresnel distance $R_F$ and $3R_F$ as a function of zenith angle given in wavelengths. The blue solid line is given by $R_{Earth}(1-R_F\cos(\theta_z)/R_{Earth})$ and the red dashed line by $R_{earth}(1-3 R_F\cos (\theta_z)/R_{earth})$. Figure~\ref{fig:area_integral} shows that above $3 R_F$ the contributions to the reflection are of order 10\%. This figure shows that the deviations from flatness are well below one wavelength for the range of interest to SWORD. These corrections may be of interest to an energy reconstruction analysis but for the purposes of simulating the experiment they are negligible.} 
\label{fig:flatness_deviation}
\end{figure*}

The size of Fresnel zones $R_F$, for wavelengths between 1~m and 10~m has a maximum of $\sim$2~km for the geometries of interest. Figure~\ref{fig:flatness_deviation} shows the deviation from flatness due to the Earth's curvature for a patch of size $R_F$ scaled with wavelength. The deviations from flatness are well below one wavelength for the range of interest to SWORD. These corrections may be of interest to an energy reconstruction analysis but for the purposes of simulating the experiment they are expected to be negligible.

\subsection{Surface Roughness}
Understanding how the reflection is affected by surface characteristics is crucial to the detectability and energy reconstruction uncertainties of the showers observed by SWORD. Surface roughness is incorporated into the reflection under the Kirchoff approximation of scalar fields, treated above, via the Rayleigh criterion for the diffraction effect of the roughness. The roughness model implemented in this study follows the self-affine fractal models of the surface of the Earth used in radar reflection analysis~\cite{shepard_1999}.

Many, if not most, topographic surfaces display fractal behavior which is not strictly self-similar, but which has a self-affinity such that the surface roughness scales with the distance over which it is evaluated~\cite{shepard_1999}. The root-mean-square (RMS) roughness $\sigma_h(L)=\sqrt{<(z-\bar{z})^2>}$, for measured heights $z$, scales as a power of the distance $L$ over which it is calculated. The dependence can be parameterized in terms of a reference scale $L_0$ and the corresponding roughness value at that scale is
\begin{equation}
\sigma_h(L) = \sigma_h(L_0)\left(\frac{L}{L_0}\right)^H,
\end{equation}
where $H$ is known as the Hurst parameter, $0 \leq H \leq 1$. Typical values for $H$ are of the order 0.5 for many topographic surfaces.

For the SWORD simulation we treat three types of surfaces: land, ice, and ocean. For land we assume a typical Hurst parameter of 0.5. Although the $\sigma_h$(1~m) can vary greatly we assume a typical value of 20~cm. As we will see, it is not expected that many SWORD events come from reflections on land due to the large amounts of anthropogenic noise present. However, for analysis of the data these parameters will have to be better estimated. For the ice we have estimated $\sigma_h$(120~m)=0.05~m and a Hurst parameter of 0.65 from in-situ measurements at Taylor Dome, Antarctica~\cite{elog_479}. An analysis, found in~\cite{elog_479},  of radarsat data~\cite{RadarSat} confirms that $H\approx0.65$ to length scales up to a kilometer. For the ocean, analysis of significant wave height from Jason-1 data~\cite{JASON} gives $\sigma_h$(2~km)$\approx 2.7$~m for the southern ocean. The SWORD simulation uses a $30^{\circ}\times30^{\circ}$ grid of $\sigma_h$(2~km) values ranging from 0.5-2.7~m and assumes a Hurst parameter of 0.5. 

For surface roughness with RMS deviation $\sigma_h(\rho)$ that is less than a wavelength, the scattered field strength upon reflection form a roughened dielectric surface is given by~\cite{shepard_1999} as
\begin{equation}
E_{rough} = E_{smooth}\exp\left(-2k^2\sigma^2_h(\rho)\cos^2\theta_z\right).
\end{equation}
In a Huygens-Fresnel application, such as the one in the previous subsection, where each surface area element is treated as re-emitting the incident radiation, we can assign this scattering amplitude to each area element, and thus include it explicitly in the integral of Equation~\ref{eqn:kirchoff}. Note here that $\sigma_h(\rho)$ is assumed to have explicit radial dependence, thus each annulus in the Kirchoff integral will have an assumed common RMS height variation, which changes with the radii of the annulus.

Including the surface roughness to the Kirchoff integral in Equation~\ref{eqn:area_integral} gives the expression
\begin{equation}
E^p_{rcv}=
E^p_{src}(\omega)
\mathcal{F}_p(\theta_z)
\frac{k}{2\pi i}
\frac{\cos\theta_z}{R_1 R_2}
\int_{-b/\cos\theta_z}^{-b/\cos\theta_z}dy
\int_{-\sqrt{b^2-y^2\cos^2\theta_z}}^{\sqrt{b^2-y^2\cos^2\theta_z}} dx
e^{-2k\sigma_h^2\left(\sqrt{x^2+y^2}\right)\cos^2\theta_z}
e^{ik|\mathbf{r}|}e^{ik|\mathbf{r}'|}
\end{equation}

The surface roughness is dominated by the Fresnel distance scale $R_F$. This means that the effective roughness contribution results in a roughness term 
\begin{equation}
G(f,R,\theta_z)\approx \exp\left[-2 k^2 \sigma^2_h(R_F) \cos^2\theta_z\right]
\label{eqn:roughness_factor}
\end{equation}

\begin{figure*}[h!]
\centering
\includegraphics[width=0.7\linewidth]{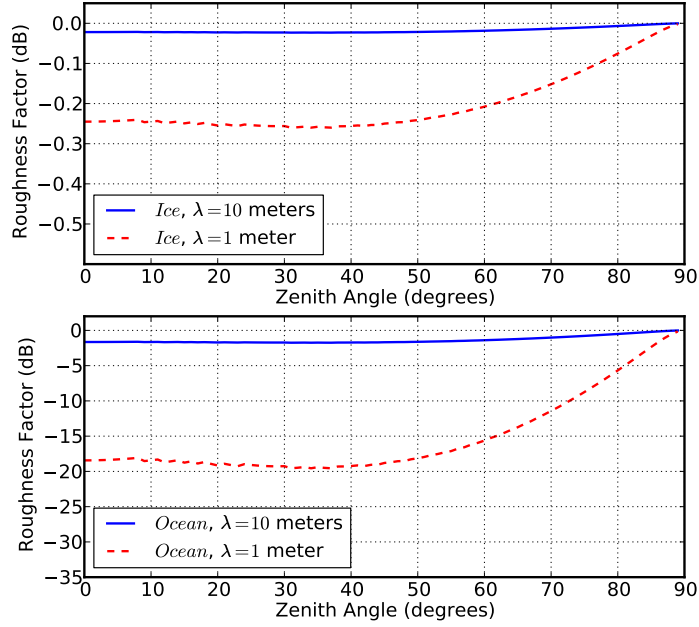}
\caption{ The roughness factor from Equation~\ref{eqn:roughness_factor} for Antarctic ice and ocean. The ice roughness results in very small attenuation factors for the VHF frequency band. For the ocean, the attenuations factor is within several dB for frequencies $<100$~MHz and becomes worsens significantly at higher frequencies.} 
\label{fig:roughness}
\end{figure*}

Examples of the attenuation due to surface roughness are given in Figure~\ref{fig:roughness}. The attenuation due to ocean surface roughness is of order a few dB for frequencies below $100$~MHz and severely affects the signal strength at higher frequencies. For ice the attenuation is not significant in any of the frequencies of interest to SWORD.

\subsection{Dispersion of VHF Pulses Through the Ionosphere}
The ionospheric plasma has a well-known dispersive effect on VHF pulses with negligible attenuation. Extensive studies of dispersion were performed with the FORTE satellite~\cite{massey_1998, roussel_2001, huang_2005, minter_2007}, on which the simulations used for SWORD are based. When a VHF pulse propagates through the ionosphere, it is dispersed from several nanosecond duration up to 10-100~$\mu$s severely affecting the signal to noise ratio. This requires that SWORD dedisperse the ionospheric effects in real-time to restore the temporal coherence of the pulse and enable efficient triggering.

There are several issues involved in developing an efficient trigger for space-based UHECR radio detection. The time and direction of incidence of the radio pulse on the detector are unknown requiring the SWORD trigger to be continuously scanning for transients. The exact state of the ionosphere is not known a priori to the degree needed for efficient triggering, which requires the dedispersion algorithm to scan over a range of ionospheric transfer function parameters. The scanning range can be constrained by knowledge and predictions of the state of the ionosphere. The optimal trigger requires a matched filter implementation but, because the search space is large and spacecraft power is limited, we need to understand what is the minimum number of assumptions and resources required to perform the dedispersion efficiently.

The implementation of ionospheric dispersion follows~\cite{moses_iono} and~\cite{myre_iono} and is described in the following. The index of refraction of the ionosphere is given by
\begin{equation}
n^2(f)=
\frac{
1-(f_p/f)^2
}
{
1-\frac{1}{2}\frac{(f_c/f)^2\sin\beta}{1-(f_p/f)}\pm\sqrt{\frac{1}{4}\frac{(f_c/f)^4\sin^4\beta}{(1-(f_p/f))^2}+(f_c/f)^2\sin^2\beta}
}
\label{eqn:index}
\end{equation}
where $f^2_p=e^2N_e/(4\pi^2\epsilon_0 m_e)$ is the plasma frequency, $f_c=eB/m_e$ is the cyclotron frequency, $N_e$ is the electron density, $e$ is the absolute value of the electron charge, $m_e$ is the electron mass, $\epsilon_0$ is the permittivity of free space, $B$ is the magnitude of the local magnetic field, and $\beta$ is the angle between the geomagnetic field vector and the direction of propagation. The $\pm$ in Equation~\ref{eqn:index} are for the ordinary and extraordinary modes of propagation The frequency dependent phase shift due to ionospheric propagation is given by
\begin{equation}
\phi(f)=\frac{2\pi}{c}\int_{T}^{R}f n(f,r) dr
\end{equation}
where $c$ is the speed of light, $T$ is the location of the transmitter, $R$ is the location of the receiver, and the integral is taken over the trajectory between $T$ and $R$.

Typical values are $f_c\approx$10~MHz and $f_p\approx5$~MHz in periods of low solar activity and $f_p\approx20$~MHz for high solar activity. For the 30-300~MHz band we can perform a Taylor expansion of the index of refraction to give
\begin{equation}
n\approx 1-\frac{1}{2}\frac{f^2_p}{f^2} + \frac{s}{2}\frac{f^2_p f^2_c}{f^3}\cos\beta -\frac{1}{4}\left[\frac{f^2_p}{2}+\left(2-\sin^2\beta\right)f^2_c\right]\frac{f_p^2}{f^4}+...
\end{equation}
where $s=\pm1$ with a positive value for wave polarized opposite to the cyclotron rotation of electrons (ordinary modes) and negative for waves polarized with the cyclotron rotation (extra-ordinary modes).

For a receiver at distance $D$ away from the transmitter, the group delay is given by
\begin{equation}
\tau(f)=\frac{D}{c}+\frac{C_2}{f^2}+s\frac{C_3}{f^3}+\frac{C_4}{f^4}
\label{eqn:group_delay}
\end{equation}
with the coefficients given by
\begin{equation}
C_2=\frac{e^2}{8\pi^2c\epsilon_0 m_e}\int_{T}^{R}N_e(r)dr,
\end{equation}
\begin{equation}
C_3=\frac{-e^3}{4\pi^2c\epsilon_0 m^2_e}\int_{T}^{R}N_e(r) B(r) \cos\beta(r) dr,
\end{equation}
and
\begin{equation}
C_4=\frac{3}{8c}\left[\frac{e^2}{4\pi^2c\epsilon_0 m_e}\int_{T}^{R}N^2_e(r) + \frac{e^4}{2\pi^2c\epsilon_0 m^3_e}\int_{T}^{R}N_e(r) B^2(r) \left((1+\cos^2\beta(r)\right) dr\right].
\end{equation}

The total electron content ($TEC$) is the vertical column density of electrons between a transmitter on the ground and a receiver a distance $h$ directly above it given by
\begin{equation}
TEC=\int_{0}^{h}N_e(z)dz.
\end{equation}
The TEC is given in units of TECU = 10$^{16}$~electrons/m$^{-2}$. We define the slanted TEC ($STEC$) for the case where the receiver is at zenith angle $\theta_z$ with respect to the transmitter on the ground at a distance $D$.
\begin{equation}
STEC=\int_{0}^{D}N_e(r)dr.
\end{equation}

For the practical implementation in the SWORD simulations we use the approximations used by Myre~\cite{myre_iono} giving
$STEC=TEC/\cos\theta_z$,
\begin{equation}
C_2=\alpha STEC.
\end{equation}
with $\alpha=e^2/(8\pi^2c\epsilon_0 m_e)=1.3445\times10^{9}$~Hz/TECU,
\begin{equation}
C_3=\frac{2m_e}{c}\alpha STEC \ B\cos\beta,
\end{equation}
and
\begin{equation}
C_4=\frac{3}{2}c^2\alpha^2 STEC^2 \frac{1}{\cos\theta_z}\left[\frac{1}{T}-\frac{\sin\theta^2_z}{h}\right],
\end{equation}
where $h$ is the receiver altitude and $T$ is the thickness of the ionosphere for a slab model.

\begin{figure*}[h!]
\centering
\includegraphics[width=0.7\linewidth]{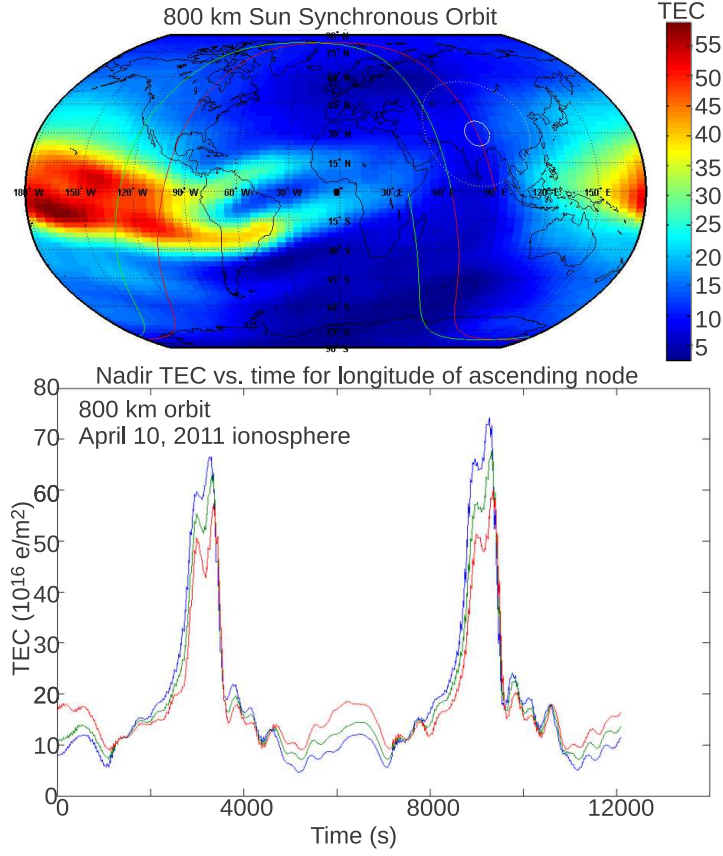}
\caption{ (Top) distribution of the ionosphere at one point in time. The distribution of total electron content (TEC) given in TECU (10$^{16}$ electrons/m$^2$) is typically worse at equatorial latitudes during the day time. An 800~km altitude, 80 degree inclination, sun synchronous orbit is superimposed on a continental outline. The small circle on the top right shows the size of the intersection of the ionosphere's peak density, at $\sim$400~km, with the field of view of the payload. The larger circle shows the contour of the payload's horizon field of view. (Bottom) the TEC distribution directly below the payload as it orbits the Earth.
} 
\label{fig:iono_distrib}
\end{figure*}

The $TEC$ values can be obtained on a daily basis from GPS derived data~\cite{ionex} and the magnetic field amplitudes and directions can be obtained from the international geomagnetic reference field (IGRF) model~\cite{igrf}. Figure~\ref{fig:iono_distrib} (top) shows the ionospheric distribution of GPS derived TEC values. The dispersion effects are expected to be best behaved near the poles and worst near the equator where the TEC has large values and fluctuations, especially when exposed to the sun. The stability of the ionosphere is dependent on solar activity meaning that for best results SWORD would need to be operate during a period away from solar maximum.

\begin{figure*}[h!]
\centering
\includegraphics[width=0.7\linewidth]{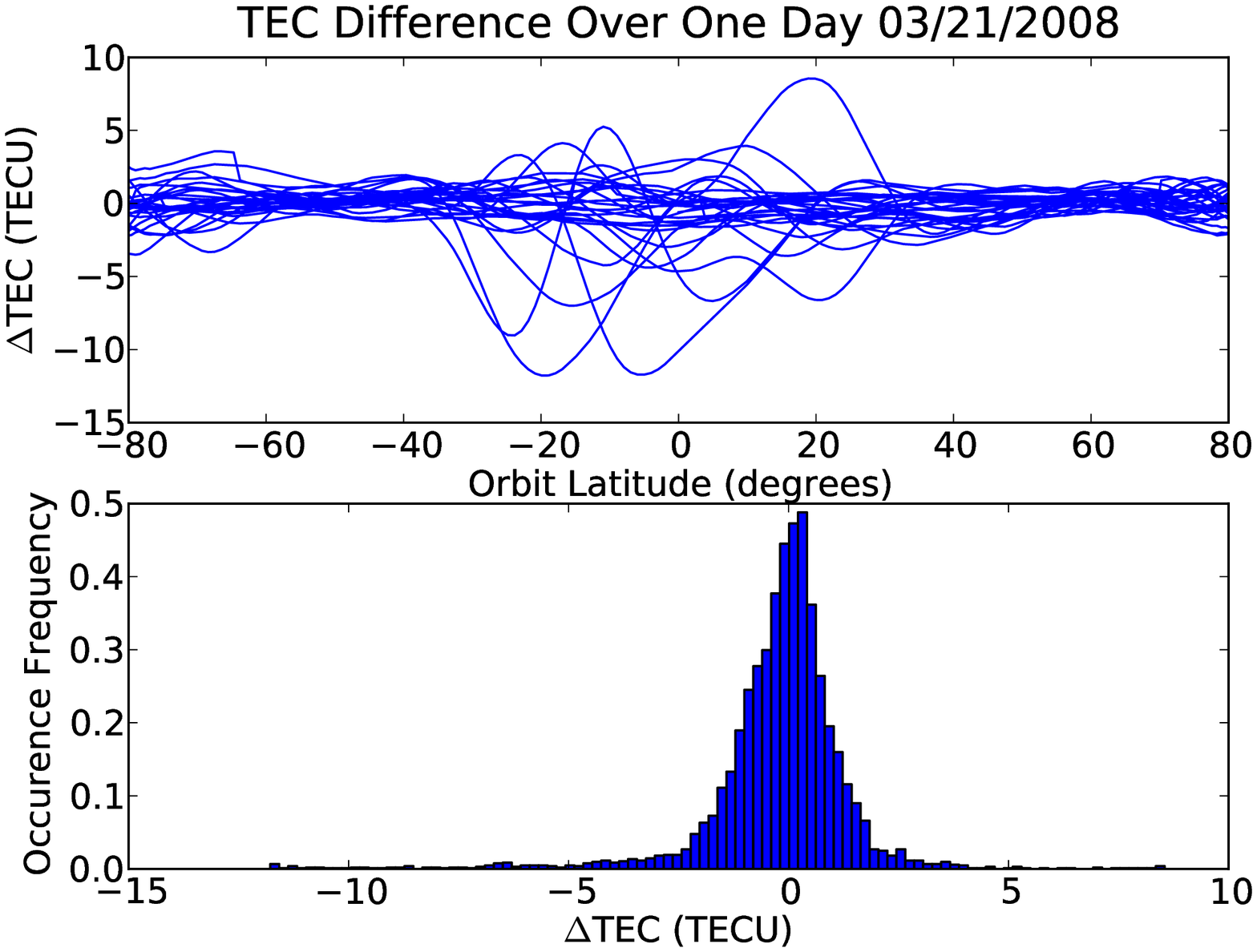}
\caption{ Difference in TEC ($\Delta$TEC)from one day to the next for a sun-synchronous orbit. Shown on top is the $\Delta$TEC directly below the payload for each of 16 orbits completed in one day as a function of orbit latitude. The histogram in the bottom is the distribution of the differences plotted above.} 
\label{fig:iono_predict}
\end{figure*}

As will be shown in Section 3, the real-time de-dispersion trigger is highly sensitive to the TEC. To save on operations count and spacecraft power it is important to constrain the TEC search space as much as possible. With the conservative assumption that the data can only be uploaded once a day, we estimated the prediction error between the ionospheric TECU over the SWORD sun-synchronous orbit by comparing the TEC values from one day to the next. Figure~\ref{fig:iono_predict} shows the difference in TEC between two consecutive days as a function of orbit latitude. The largest deviations are near the equator, as expected. The overall RMS of the difference is $\approx$2~TECU as shown in the histogram in the bottom of Figure~\ref{fig:iono_predict}. A study with the FORTE satellite, comparing the reconstructed TEC using the observed dispersion of VHF wideband pulses, compared to MAGIC, which is a GPS derived ionospheric solution, shows a scatter of $\approx$2.8~TECU~\cite{huang_2005}. Altogether, it is expected we can bound the ionospheric TEC in the field of view of SWORD by about 3-4~TECU. 

\subsection{Noise Backgrounds}
The dominant source of noise in the VHF band is the galactic noise background with an average value of $\sim10^4$~$K$ at 30~MHz decreasing as a power law with index -2.8 to $\sim10$~$K$ at 300~MHz. To estimate the noise on SWORD we use the galactic background maps provided by ~\cite{oliveira_2008}. The SWORD antenna, which will be described later, has a 7~dBi gain. The noise maps of ~\cite{oliveira_2008} are therefore averaged in $30^{\circ}\times30^{\circ}$ bins on the sky. This averaged galactic noise background can vary by about a factor of 8 depending on whether the galactic center is in the field of view. 

The SWORD antennas are pointed towards the Earth's horizon. This means that a portion of the galactic noise contribution is reflected off the surface of the Earth prior to exciting the antenna. For an antenna with gain $g(f,\theta,\phi)$, the effective antenna temperature is given by
\begin{equation}
T_A(f)=\frac{\int_{\Omega} d\Omega \ g(f,\theta,\phi) T_{source}(f,\theta,\phi)}
{\int_{\Omega}  d\Omega \ g(f,\theta,\phi)}
\end{equation}
For the SWORD geometry, the noise sources are the direct observation of the sky noise temperature $T_{sky}$, which consists of about 75\% of the antenna beam pattern, and the observation of the ground, which consists of the remaining 25\% of the beam pattern. The ground thermal noise is a combination of the reflected sky noise radiation, which is weighted by the Fresnel reflection coefficient $\mathcal{F}^2T_{sky}$ and thermal emission from the surface $(1-\mathcal{F}^2)T_{surface}$. The surface noise temperature is approximated at $240^{\circ}$ for ice and $270^{\circ}$ for land. The ocean is assumed to be completely reflective at VHF frequencies.

The anthropogenic noise background is also modeled in the SWORD simulation. The FORTE satellite made extensive studies of the radio noise backgrounds on Earth~\cite{burr_2004, burr_2005}. For the SWORD simulation, we summarize the effects of anthropogenic noise with a noise factor $NF$, which is a proportionality factor to the antenna noise temperature due to the sky noise and inversely proportional to frequency 
\begin{equation}
T_{anthro}(f)=T_{ant}(f) (NF-1) \left(\frac{39\mbox{ MHz}}{f}\right)
\end{equation}
The noise factor $NF$ is strongest over North America, Europe, and East Asia, peaking at $NF\sim40$, and weakest over open ocean, where there is practically no contribution.

The final source of thermal noise is the system temperature $T_{sys}$ of the instrument itself. A typical value of $T_{sys}=90$~K is used based on current amplifier technology.


\section{Mission Architecture}
\begin{figure*}[h!]
\centering
\includegraphics[width=0.7\linewidth]{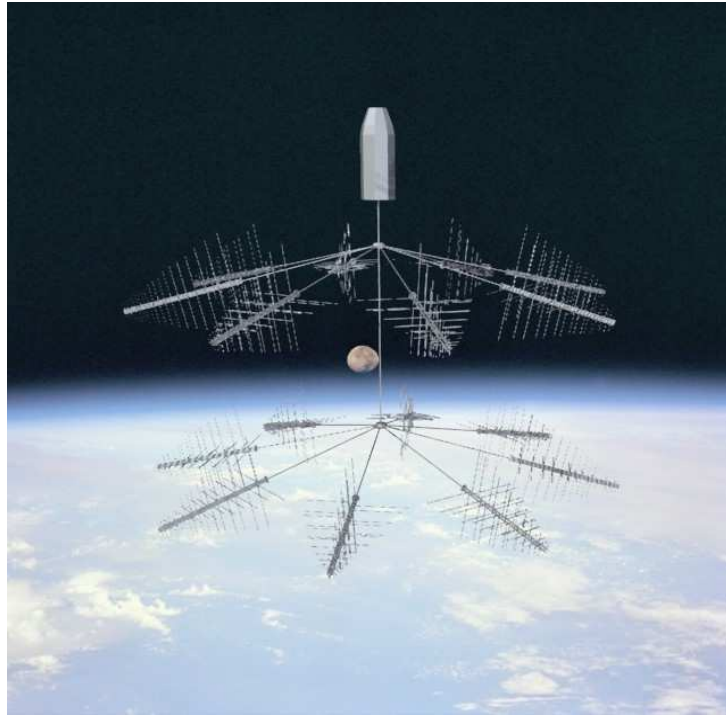}
\caption{ A 16 antenna version of the SWORD spacecraft. Two rings of 8 antennas each deploy from a Pegasus Fairing envelope. The cylindrically symmetric configuration provides synoptic observation of the Earth's surface, which provides a full-sky survey of UHECRs as well as an added handle for identifying radio frequency interference.} 
\label{fig:sword_image}
\end{figure*}

The SWORD payload consists of an array of dual-polarized log-periodic dipole array antennas. Each antenna polarization has its own signal chain that is fed to a trigger system and data recording system. The payload can fit in a Pegasus Fairing envelope with 1.94~meter height and diameter that tapers from 1.16 - 0.72~meters. A version of the SWORD payload with 16 antennas is shown in Figure~\ref{fig:sword_image}. The orbit of choice is an 80$^{\circ}$ inclination, sun-synchronous orbit, with altitude ranging from 600-800~km. The high inclination is driven by observation of the poles where the ice provides a relatively smooth reflector with generally radio quiet locations. Placing the orbit at the terminator provides enough exposure to the sun to power the spacecraft while keeping away from the most excited portions of the ionosphere. 

\begin{figure*}[h!]
\centering
\includegraphics[width=\linewidth]{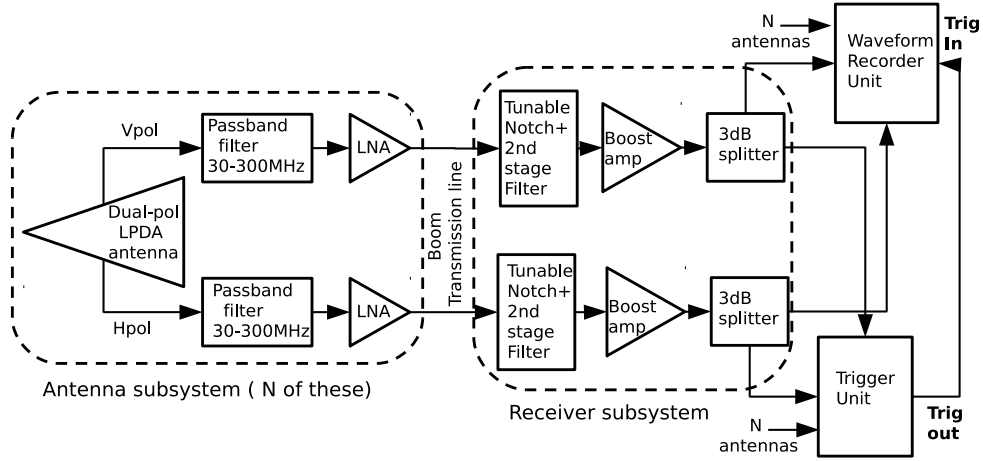}
\caption{The SWORD system diagram. Each antenna polarization has its own signal chain consisting of passband filters and low noise amplifiers. The signals are fed into the central instrument with tunable notch filters for mitigation of carrier wave interference. The second stage filter prevents any additional out of band noise picked up by the signal chain from being aliased. After a second stage of amplification the antenna signals are split into a trigger unit and a digital recording unit. } 
\label{fig:sword_system_diagram}
\end{figure*}

The spacecraft will be equipped with a GPS receiver as well as attitude sensors for tilt and heading. The SWORD system diagram is shown in Figure~\ref{fig:sword_system_diagram}. An additional processor will be used to discriminate against lightning and anthropogenic backgrounds. For signals digitized at 600~Msa/s with 3-bit resolution and a record length of $\lesssim$100~$\mu s$ in a payload consisting between 6-16 dual-polarized antennas, the size of a raw event ranges between 2.2-5.7 Mbits. Based on satellite maps of lightning strike rates~\cite{noaa_lightning}, lightning is expected to trigger the payload at average rates of $\sim$1~Hz. Keeping the telemetry rates down to several Gbit/day will require efficient filtering of background events. Fortunately, it is well known that the UHECR signature is directly correlated to the direction of the geomagnetic field. Along with wide-band spectral and event isolation requirements, it is expected that the false positives rate will be low.

\subsection{Antenna}
The spatial constraints of the Pegasus envelope and VHF wavelengths, ranging from 1-10 meters, requires that the SWORD antennas be deployable. The wire structure of the log-periodic dipole antenna naturally lends itself to this application. Deployable LPDA antennas have been flown on other missions, most notably FORTE.

\begin{figure*}[h!]
\centering
\includegraphics[width=\linewidth]{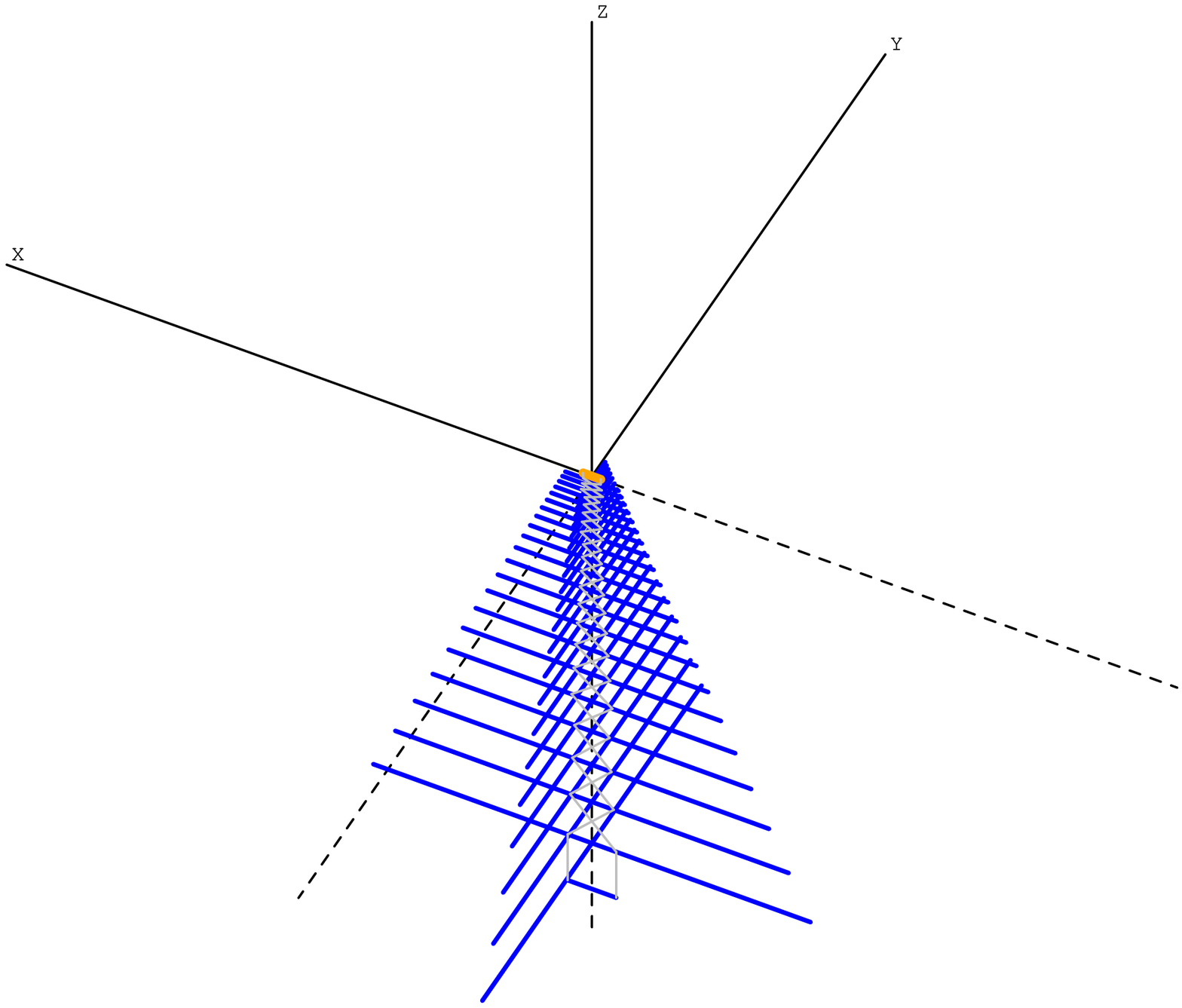}
\caption{The SWORD 30-300~MHz log-periodic array (LPDA) antenna. The boom length is 4.5 meters long with longest dipole span of 3.7 meters. The wire structure of the LPDA lends itself to a deployable structure.} 
\label{fig:sword_lpda}
\end{figure*}

The design shown in Figure~\ref{fig:sword_lpda} provides a radiation pattern with gain ranging form 7-5 dBi in the frequencies of interest. The HPBW of the beam pattern ranges from 100$^{\circ}$-120$^{\circ}$ providing a wide field of view for UHECR detection.

\subsection{Trigger}
The trigger system uses the ANITA-1~\cite{gorham_2009a,gorham_2009b} and ANITA-2~\cite{gorham_2010} threshold and coincidence approach. The signal is split into bands to protect the system against RF interference. This way, bands can be masked if they are being saturated. Each band triggers on a low threshold at MHz rates with temporal coincidence and geometric requirements to bring the rate down to a 5-10~Hz payload trigger rate. In the case of SWORD, the individual bands have to be digitized in order to dedisperse the effect of the ionosphere in real-time. 

The current trigger model uses 7 bands for each polarization with bandwidth distributed according to the the inverse frequency phase delay. The band limits are given by the recurring relation $f_{i+1}=f_i \Delta$ where $\Delta=\left(300 MHz/30 MHz\right)^{1/7}$.
\begin{table}
\centering
\begin{tabular}{|c|c|c|}
\hline
$f_{Low}$ (MHz) & $f_{High}$ (MHz) & Bandwidth (MHz) \\
\hline
\hline
30 & 42 & 12 \\
\hline
42 & 58 & 16 \\
\hline
58 & 80 & 22 \\
\hline
80 & 112 & 32 \\
\hline
112 & 155 & 43 \\
\hline
155& 216 & 61 \\
\hline
216 & 300 & 84 \\
\hline
\end{tabular}
\caption{The SWORD trigger bands distributed according to inverse frequency phase delay. See text for details.}
\end{table}

The trigger requires that at least 4 of the 14 total bands (7 for each polarization) exceed a threshold of $1.6 \sigma$ where $\sigma$ is the voltage RMS of the noise. In addition, the full band signal has to exceed $4.2\sigma$. The threshold crossing accidentals rate is given by the cumulative distribution function of the Rician distribution for noise $p=\exp[-v^2_{thresh}/(2\sigma^2)]$. For a digitized signal, accidentals rate is given by the sampling rate multiplied by the probability of a threshold crossing. The threshold crossing accidentals trigger rate is estimated to be $r_{sb}\approx170$~MHz for each sub-band while the full-band accidentals rate is $r_{fb}\approx88$~kHz. The probability that 4 out of 14 bands trigger is approximated by $P\approx ^{14}C_4 (r_{sb} \tau)^4(1-r\tau)^{10}$~\cite{gorham_2009a} where $\tau$ is the trigger window. A $\tau=3$~ns window gives $P\approx0.05 $. Combined with the probability that the full band is above 4.2$\sigma$ ($p_{fb}=2.6\times10^{-4}$) gives a single antenna accidental probability of $1.3\times10^{-5}$. At 600~Msa/s the total accidentals rate per antenna is $\approx 1.1$~Hz. For the full set of $N_A$ antennas, without geometric coincidence requirements, the total payload accidentals rate is given by $N_A\times 1.1$~Hz.

Prior to applying the threshold and coincidence trigger, the signals need to be dedispersed. The algorithm applies coherent dedispersion via cross-correlation with a template. The dispersion phase, is given by
\begin{equation}
\phi = 2\pi\frac{C_2}{f}+s2\pi\frac{C_3}{f^2}
\end{equation}
where $C_2$ and $C_3$ are the coefficients for the group delay ($\tau=-\partial\phi/\partial\omega$) in Equation~\ref{eqn:group_delay}. The digitized voltage data for a given band is cross-correlated with the dispersion phase, with the opposite sign, to dedisperse the signal. The birefringence of the ionosphere acts on circularly polarized modes where the SWORD antenna observes linearly polarized modes. The template for each polarization is therefore derived starting from the assumption that the signal has polarization angle $\theta_p$ along $\mathbf{B}_{\perp}$ as defined in Equation~\ref{eq:geo_synch_param}. Let $\mathbf{\hat{H}}$ and $\mathbf{\hat{V}}$ define the horizontal and vertical polarization vector directions. The Fourier coefficient of the template is given by  $\mathbf{W} = e^{i\omega t_0}\left(\cos\theta_p \mathbf{\hat{H}} + \sin\theta_p \mathbf{\hat{V}}\right)$. 
The transformations between right/left circular and horizontal/vertical polarizations are
\begin{equation}
R=(H+iV)/\sqrt{2}
\end{equation}
\begin{equation}
L=(H-iV)/\sqrt{2}
\end{equation}
\begin{equation}
H=(L+R)/\sqrt{2}
\end{equation}
\begin{equation}
V=i(L-R)/\sqrt{2}
\end{equation}

\begin{figure*}[h!]
\centering
\includegraphics[width=0.7\linewidth]{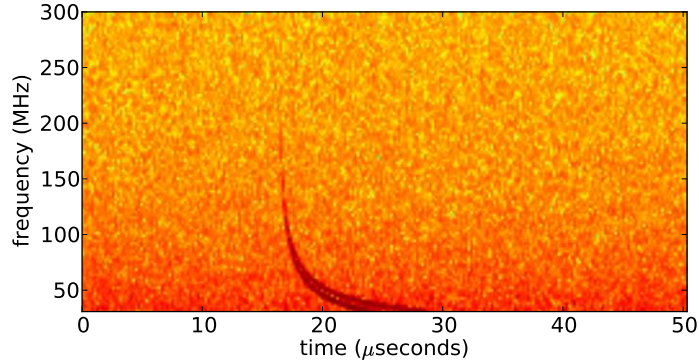}
\caption{Simulated cosmic ray event with $2\times10^{20}$~eV propagated through an ionosphere with STEC$\approx$9 TECU. The splitting at lower frequencies is due to the geomagnetic field. Although this pulse is strong enough to be observed above the noise background in this spectrogram, pulses at lower energies are buried in the noise. It is only after coherent dedispersion that the power at each band will be temporally coherent resulting in a pulse that can be distinguished from the noise background.} 
\label{fig:spectrogram}
\end{figure*}

The right and left circularly polarized modes of the pulse prior to dispersion are given by
\begin{equation}
W_R = \frac{e^{i\omega t_0}e^{i\theta_p}}{\sqrt{2}}
\end{equation}
\begin{equation}
W_L = \frac{e^{i\omega t_0}e^{-i\theta_p}}{\sqrt{2}}
\end{equation}
After dispersion through the ionosphere, $W_R$ and $W_L$ will carry additional dispersion phase.
\begin{equation}
W_R = \frac{e^{i\omega t_0}e^{i\theta_p}e^{i\phi_R}}{\sqrt{2}}
\end{equation}
\begin{equation}
W_L = \frac{e^{i\omega t_0}e^{-i\theta_p}e^{i\phi_L}}{\sqrt{2}}
\end{equation}
Transforming back to $H$ and $V$ results in 
\begin{equation}
W_H = e^{i\omega t_0}\frac{e^{i(\phi_R+\theta_p)}+e^{i(\phi_L-\theta_p)}}{2}
\end{equation}
\begin{equation}
W_V = e^{i\omega t_0}\frac{e^{i(\phi_R+\theta_p)}-e^{i(\phi_L-\theta_p)}}{2i}
\end{equation}
An example of the birefringence of the ionosphere, shown in Figure~\ref{fig:spectrogram}, is due to the combination of left and right dispersion modes.

The dedispersion algorithm proceeds to cross-correlate the data with the template
\begin{equation}
U_{Vpol}= V_{Vpol}\otimes W_V
\end{equation}
\begin{equation}
U_{Hpol}= V_{Hpol}\otimes W_H
\end{equation}

The parameters to which this algorithm is sensitive to are the TEC precision with which the dispersion can be modeled, the amplitude and direction of the geomagnetic field, and additional dispersion due to higher order terms in the expansion of the dispersion phase. The results of the trigger sensitivity to SWORD for these parameters is shown in Section 4.

\subsection{On-board Data Filtering}
Limited telemetry rates require that SWORD have an efficient way to analyze data on board. Anthropogenic backgrounds are a major concern as they are likely to trigger the payload at a high rate. A spatio-temporal cluster rejection algorithm will need to be run on board similar to those successfully employed in the analyses of both ANITA flights~\cite{gorham_2009b,gorham_2009a,gorham_2010}. Strong anthropogenic backgrounds cover $\sim$30\% of the surface of the Earth. As will be shown in the next section, although the sensitivity of 10$^{19}$-10$^{20}$~eV is likely to be severely affected, events with energies $>10^{20}$~eV, which are most relevant for charged particle astronomy, are bright enough so that detectability will not be as strongly affected.

In the remaining $~70\%$ of the surface of the Earth, mostly covered by ocean, lightning is a major concern. World lightning frequency maps~\cite{noaa_lightning} show lightning occurrence rates of 0.1-0.4 events per km$^2$ per yr over the souther ocean. For the SWORD payload, at an altitude of 800~km, with a visible area of 22 million km$^2$, this translates to a rate of $\sim$0.5~Hz. Although this seems like a low rate, assuming 3-bit digitizers running at 600 Msa/s with a record length of 100~$\mu$s (to accommodate for ionospheric dispersion) produces a data rate of 16~Gbits per antenna per day. With 12 antennas this is 190~Gbit/day. Data compression prior to transmission will be of aid but to keep telemetry rates reasonable the lightning false positives rate in the on-board analysis will have to be $\lesssim$10\%. 

The main discriminator between lightning and geosynchrotron pulses is the requirement that the polarization of the UHECR be perpendicular to the geomagnetic field. Lightning is strong enough to pass the trigger thresholds on intensity alone but on-board analysis will be able to determine the polarization of the signal. The lightning pulse is also of longer duration than the UHECR and it is expected that better discriminators can be developed with further applications of the measurements from FORTE~\cite{jacobson_1999} and the upcoming UHECR events from the third flight of ANITA~\cite{hoover_2010}.


\section{Expected Performance}
\begin{figure*}[h!]
\centering
\includegraphics[width=0.7\linewidth]{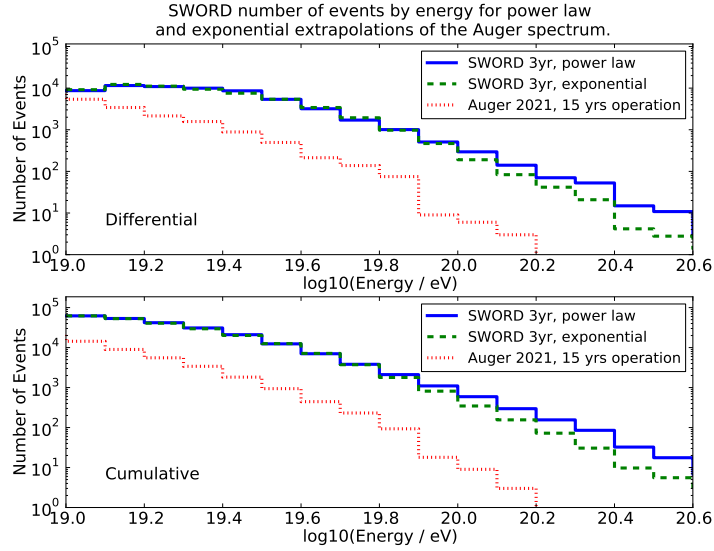}
\caption{The number of events expected with SWORD assuming a 2018 launch and 3 years of operation. The top histogram shows the differential distribution for SWORD for both a power law and exponential extrapolation of the Auger spectrum. The event count spectrum for Auger, at the time of completion of the SWORD, is shown for comparison. The bottom histogram shows the expected cumulative event counts.} 
\label{fig:sword_num_events}
\end{figure*}

\begin{figure*}[h!]
\centering
\includegraphics[width=0.7\linewidth]{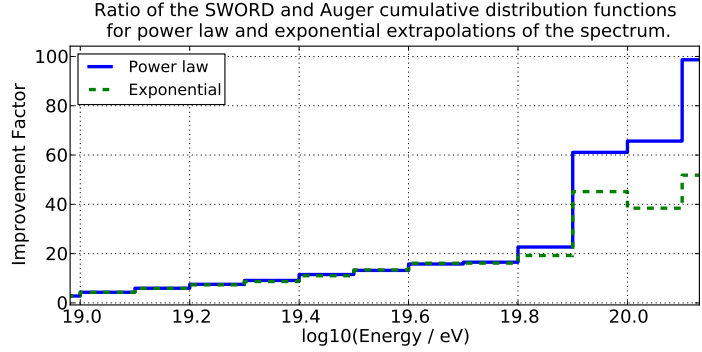}
\caption{The improvement factor for the cumulative distributions shown in Figure~\ref{fig:sword_num_events} expected from SWORD over Auger assuming a 2018 launch and 3 years of operation. At super-GZK energies the improvement factor ranges from 40 to 70.} %
\label{fig:improvement_factor}
\end{figure*}

\begin{figure*}[h!]
\centering
\includegraphics[width=0.7\linewidth]{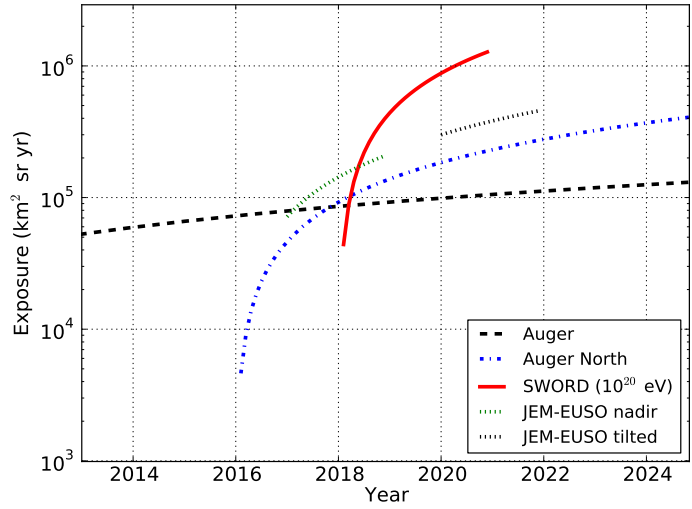}
\caption{The SWORD exposure compared to the Auger and JEM-EUSO values from~\cite{Kotera_2011}. The SWORD values shown here is for events with energy $>10^{20}$~eV for a 12 antenna version of SWORD. Unlike ground arrays, the exposure of SWORD is strongly energy dependent and rises steeply with increasing energy.} 
\label{fig:exposure}
\end{figure*}

The expected number of events accumulated by SWORD with a 3 year mission is shown in Figure~\ref{fig:sword_num_events}. At energies beyond $3\times10^{19}$ the spectrum is extrapolated by either an exponential fall-off or a broken power law. The improvement factor over Auger, assuming a 2018 launch and 3 year flight, ranges between factors of 40 to 70 at extremely high energies (see Figure~\ref{fig:improvement_factor}). The SWORD exposure for $10^{20}$~eV, compared to Auger and JEM-EUSO values from ~\cite{Kotera_2011} is shown in Figure~\ref{fig:exposure}.

\begin{figure}[h!]
\centering
\includegraphics[width=0.7\linewidth]{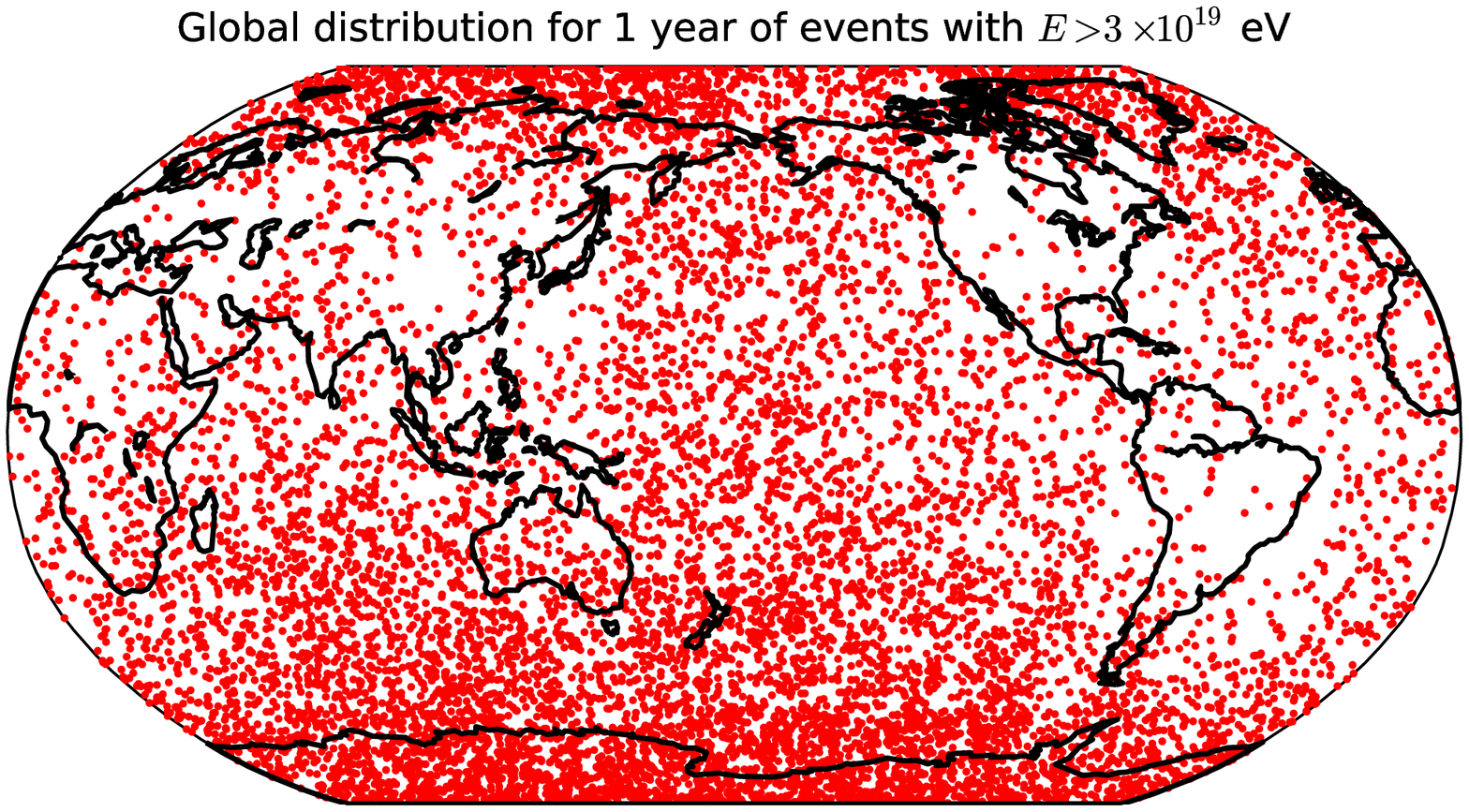}
\caption{Global distribution for one year's worth of events with energy greater than $3\times10^{19}$~eV. The events are concentrated on the ice caps primarily due to the low surface roughness factors of ice (see Figure~\ref{fig:roughness}). The oceans are highly reflective but surface roughness is significantly larger, particularly at the higher end of the SWORD spectrum ($>100$~MHz). The low density of events over land is primarily due to increased anthropogenic backgrounds. }
\label{fig:map_3e19}
\end{figure}

\begin{figure}[h!]
\centering
\includegraphics[width=0.7\linewidth]{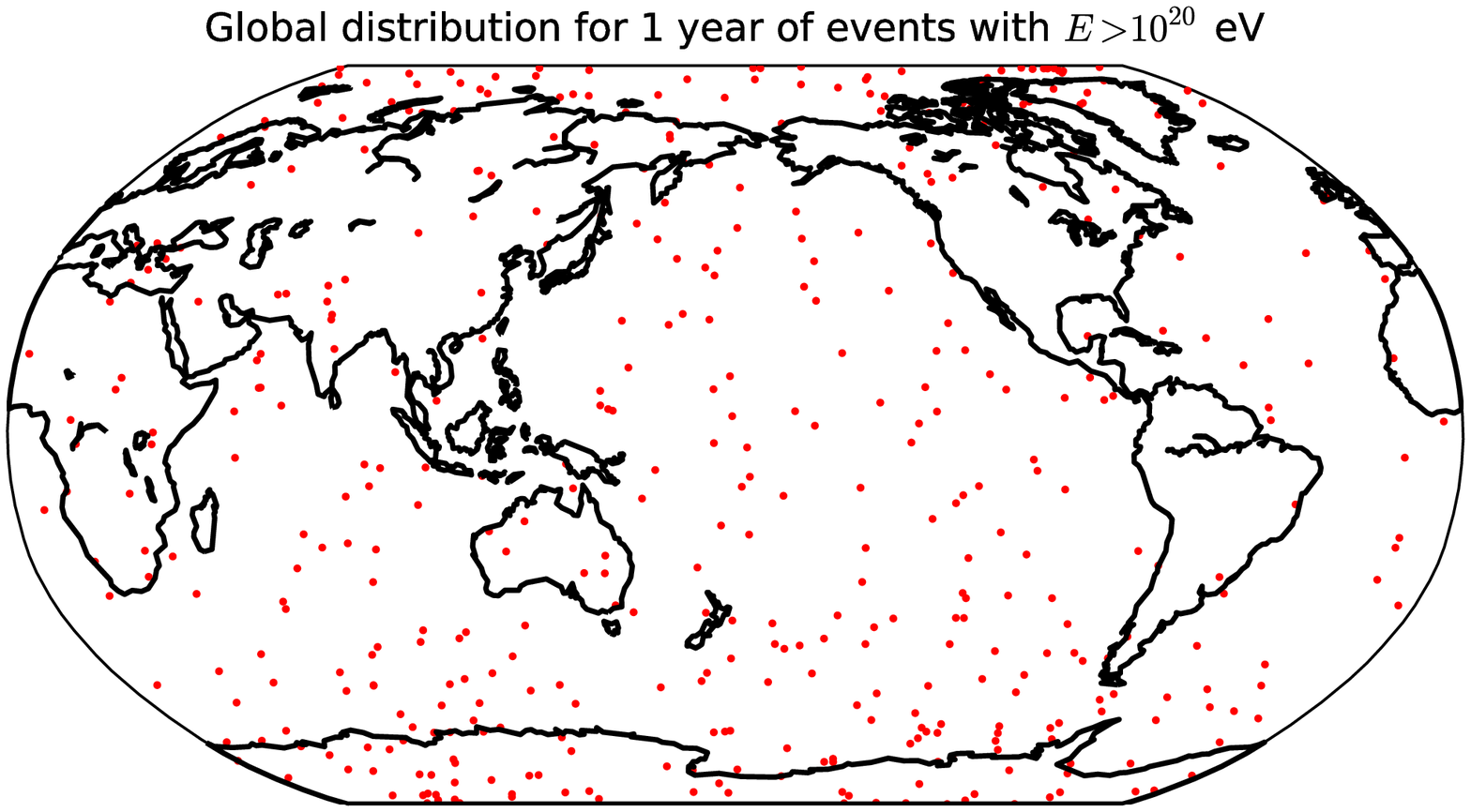}
\caption{Global distribution for one year's worth of events with energy greater than $10^{20}$~eV. Due to the large signal strength provided at these energies, the distribution of triggered events on the globe is not significantly affected by surface roughness or anthropogenic backgrounds.} 
\label{fig:map_1e20}
\end{figure}

The distribution of one year's worth of events are shown in Figure~\ref{fig:map_3e19} for energies $>3\times10^{19}$~eV and in Figure~\ref{fig:map_1e20} for energies $>10^{20}$~eV. The absence of events on the continents is primarily due to anthropogenic noise. The concentration of events on the Antarctic ice cap with respect to the ocean is due to the smoother reflection surface of the continent.

\begin{figure}[h!]
\centering
\includegraphics[width=0.7\linewidth]{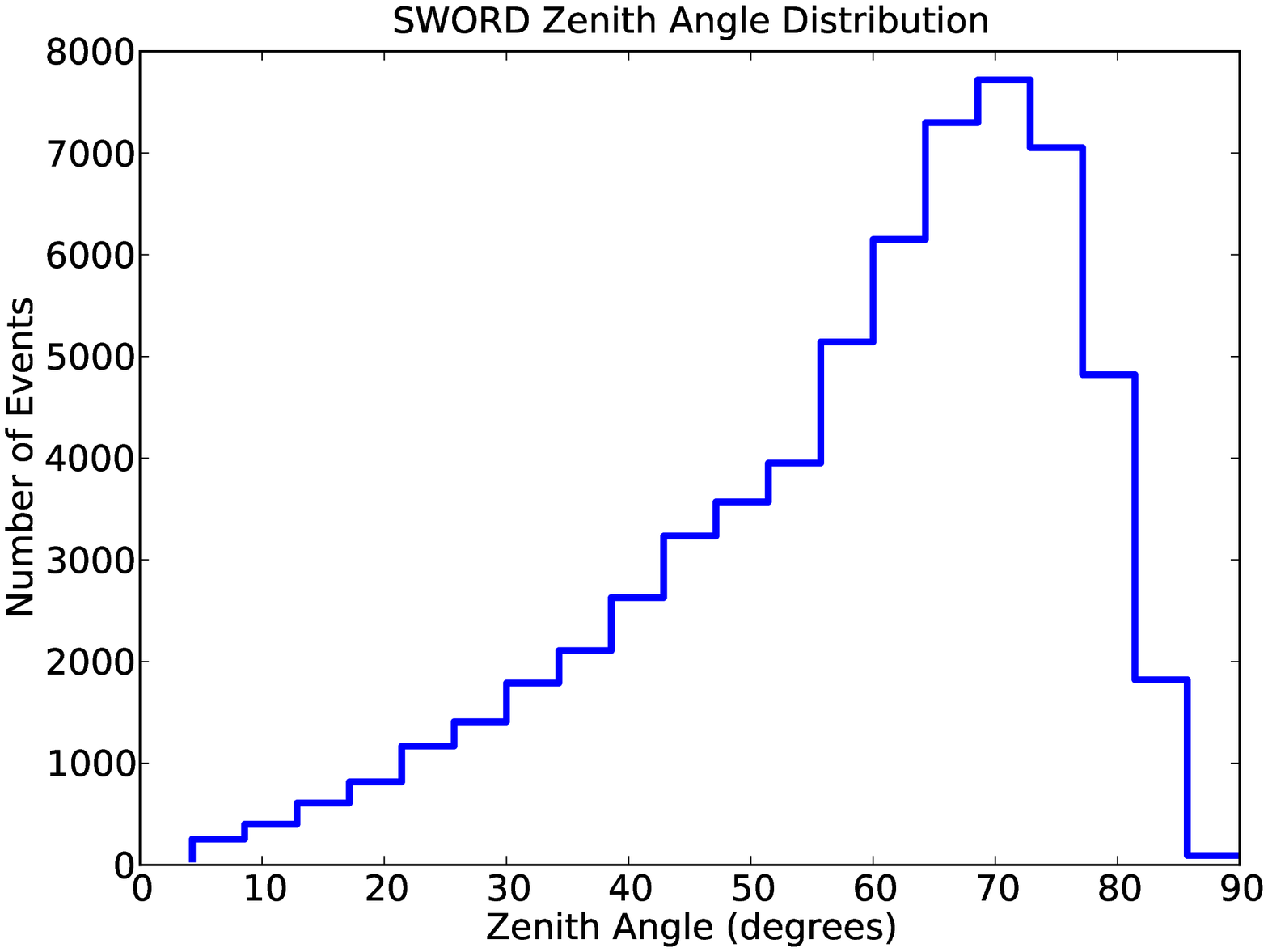}
\caption{Distribution of zenith angles observed by SWORD. The distribution peaks at around 70 degrees where the exposure is high but the signal is not attenuated by the obliquity factor. The decrease in the number of events at lower zenith angles is due to the reduced collection area corresponding to those directions.} 
\label{fig:zenith_ang_distrib}
\end{figure}

The distribution of zenith angles for SWORD triggered events is shown in Figure~\ref{fig:zenith_ang_distrib}. At higher zenith angles, the SWORD payload has a larger collection area. However, the signal is attenuated at high zenith angles due to the obliquity factor in Equation~\ref{eq:geo_synch_param}. At lower zenith angles the collection area decreases resulting in a corresponding decrease in the number of events.

\begin{figure}[h!]
\centering
\includegraphics[width=0.7\linewidth]{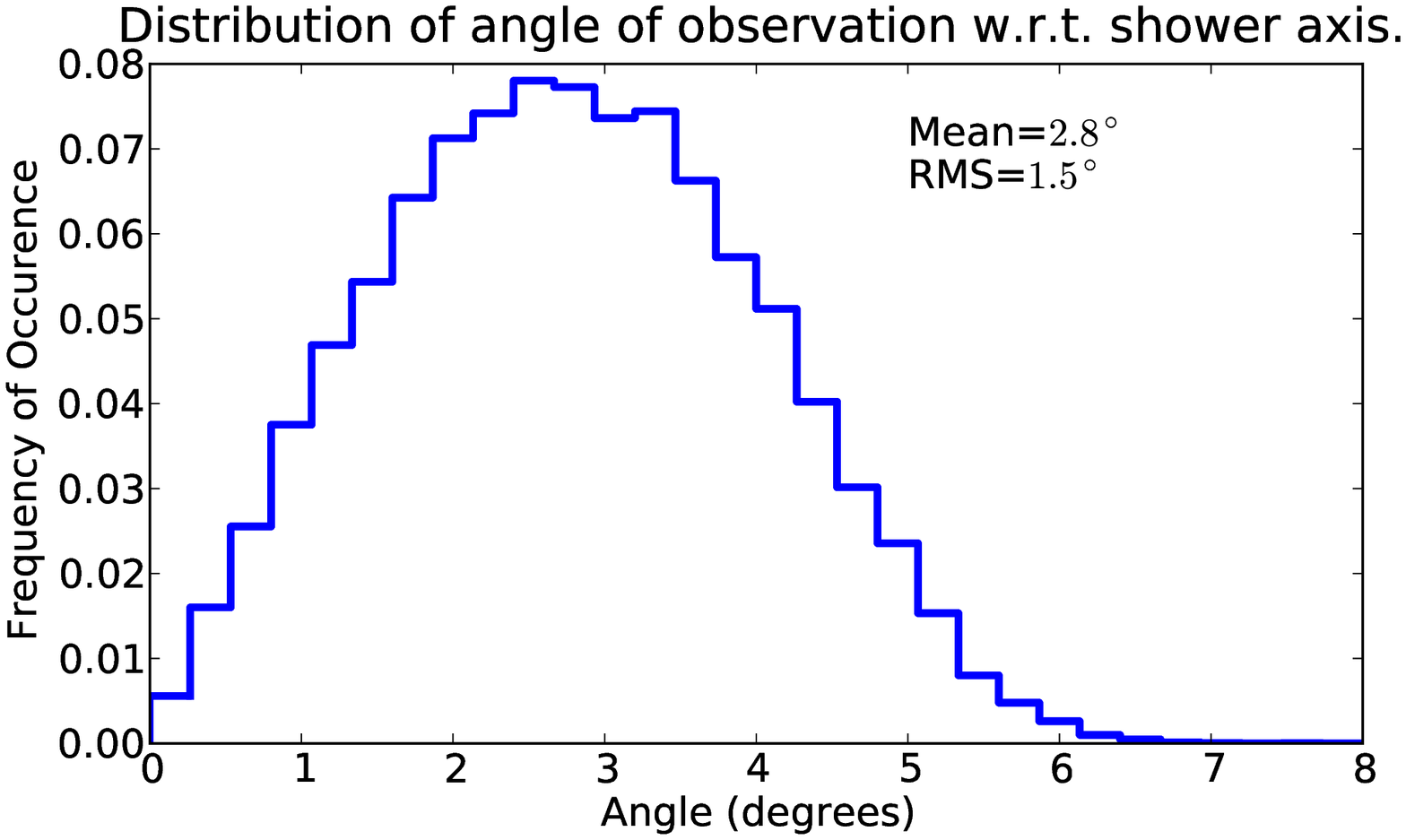}
\caption{Distribution of observation angles with respect to the shower axis for the expected ensemble of triggered SWORD events. The mean value is 2.8 degrees with a root mean square of 1.5 degrees. The energy dependence of this angle tends to be wide for higher energies, where there is enough power the trigger away from the shower axis, and narrower for low energies, where the observation needs to be near the shower axis for the power to be above the noise.} 
\label{fig:triggered_angular_distrib}
\end{figure}

Although an antenna array can reconstruct the direction of an incident plane wave with relatively high accuracy, as was demonstrated with ANITA~\cite{gorham_2009a}, the ability to reconstruct the direction of the UHECR event is limited by the unknown angle of observation with respect to the shower axis. The distribution of angles with respect to the shower axis for triggered events is shown if Figure~\ref{fig:triggered_angular_distrib}.  The mean angle is $2.8^{\circ}$ with a RMS of $1.5^{\circ}$. It is expected that the interferometric direction reconstruction of the data will be significantly better than this so this distribution remains the dominant source of error for directional reconstruction. It is possible that radio spectral analysis could yield improved event direction reconstruction with application of techniques similar to~\cite{belov_2012}. However, estimating the angle of observation with respect to shower axis using this technique does require that the attenuation due to surface roughness, which has a similar effect, be well understood. This direction is worth further exploration but we take the conservative approach here and not include this in our estimates.  The technique in~\cite{belov_2012} is currently being explored in simulations for application to the ANITA cosmic ray data expected at the beginning of 2014.

The energy reconstruction is obtained by dividing the recorded voltage $v_f$ at each frequency $f$ by 
\begin{equation}
E_f=(10^{19}\mbox{ eV})\frac{v_f}{h_{eff}(f)} \left(S(f)\frac{\mathbf{B_{\perp}}}{B_{ref}}\cos\theta_z\mathcal{F}(\theta_z,f)\right)^{-1}
\end{equation}
where the direction reconstruction provides $\theta_z$ and the geomagnetic field. The Fresnel coefficient $\mathcal{F}(\theta_z,f)$ can be estimated based on the surface of reflection (ice or ocean for most events). This estimate does not include surface roughness effects or attenuation due to observing the UHECR signal off the shower axis.

The total energy is estimated by a weighted average across the bands according to 
\begin{equation}
E_{measured}=\frac{ \sum_{f} E_f (v_f/n_f)^2} {\sum_{f} (v_f/n_f)^2} 
\end{equation}
where $n_f$ is the noise RMS voltage.

\begin{figure}[h!]
\centering
\includegraphics[width=0.7\linewidth]{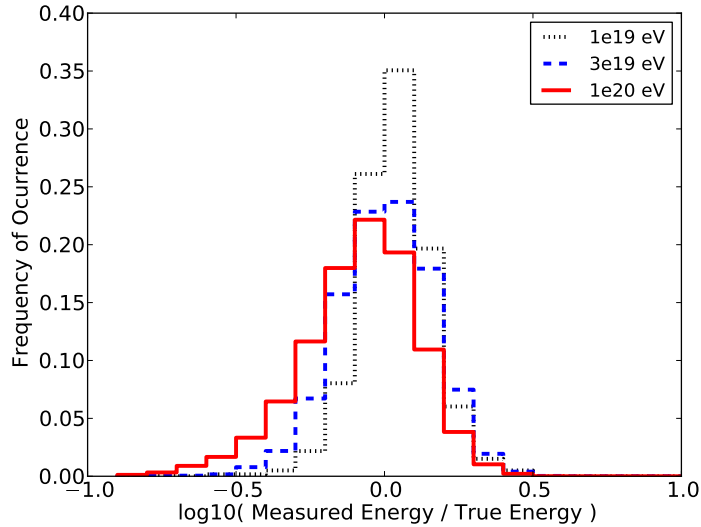}
\caption{Distributions of the ratio of measured energy to true energy for $10^{19}$, $3\times10^{19}$, and $10^{20}$~eV UHECRs. The width of the distribution is due to the combined effect of noise backgrounds and the distribution of angle of observation with respect to the shower axis, which is not accounted for in the reconstruction.} 
\label{fig:energy_recon_distrib}
\end{figure}

\begin{figure}[h!]
\centering
\includegraphics[width=0.7\linewidth]{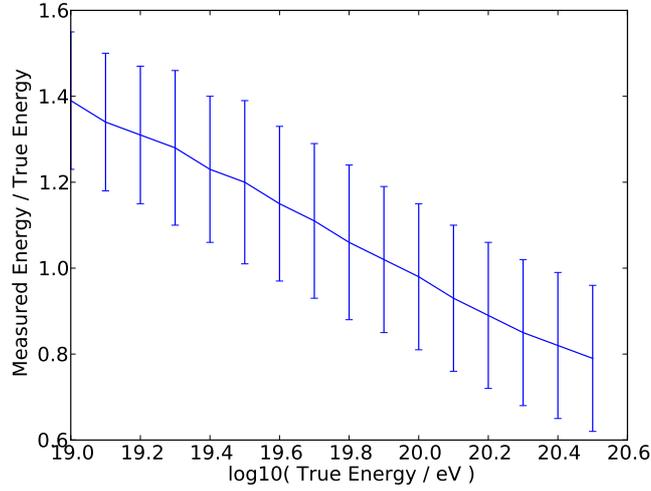}
\caption{Statistics of the ratio of measured energy to true energy. The mean values are error bars are derived from distributions such as those shown in Figure~\ref{fig:energy_recon_distrib}.} 
\label{fig:energy_recon_stats}
\end{figure}

The distribution of reconstructed vs. true energy is shown in Figure~\ref{fig:energy_recon_distrib}. The increasing asymmetry at higher energies is due to the fact that the UHECR radiation strength increases allowing the payload to trigger at larger angles off shower axis. In other words, the distribution shown in Figure~\ref{fig:triggered_angular_distrib} is wider at higher energies and narrower at lower energies. Since we are not assuming that the angle of observation with respect to the shower axis can be known, it becomes a dominant source of uncertainty in the energy reconstruction as well as the direction reconstruction of events. The statistics of energy reconstruction are shown in Figure~\ref{fig:energy_recon_stats}. The bias and errors are stable and well understood making a statistical energy reconstruction possible.

The results shown so far do not take into account the dispersion effects of the ionosphere. In the VHF band the ionosphere does not significantly attenuate radio signals and the dispersion effects are understood well enough so that they can be removed in analysis. The greatest impact is on triggering efficiency. The ability of the SWORD mission to match the results shown in this paper depends on the real-time de-dispersion capabilities of the payload, which are limited by operations count and spacecraft power consumption. A full time-domain simulation of the dedispersion algorithm, using matched filtering for the ionospheric TEC and the geomagnetic field parameters was developed to guide the hardware requirements. 

\begin{figure}[h!]
\centering
\includegraphics[width=0.7\linewidth]{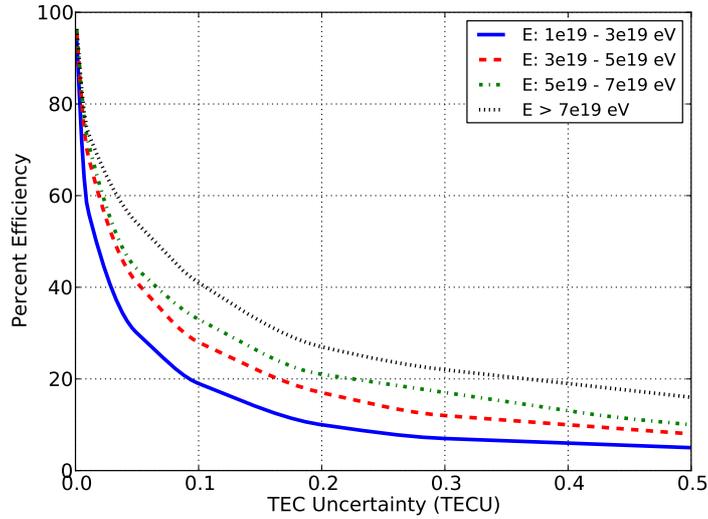}
\caption{Efficiency of real-time dedispersion trigger, relative to the trigger described in section 3, assuming cross-correlation template parameter errors in the TEC parameter. The SWORD trigger produces waveform templates based on a value for TEC and the geomagnetic field value. As the TEC value deviates from the true TEC, the efficiency drops.} 
\label{fig:tec_efficiency}
\end{figure}

The results of the simulations showing the trigger inefficiencies given the precision with which the ionospheric TEC is sampled is shown in Figure~\ref{fig:tec_efficiency}. To retain 50\% efficiency at $5\times10^{19}$~eV the dedispersion algorithm is required to sample templates in 0.05~TECU steps where the ionosphere can be constrained to within a few TECU (see Figure~\ref{fig:iono_predict}).

\begin{figure}[h!]
\centering
\includegraphics[width=0.7\linewidth]{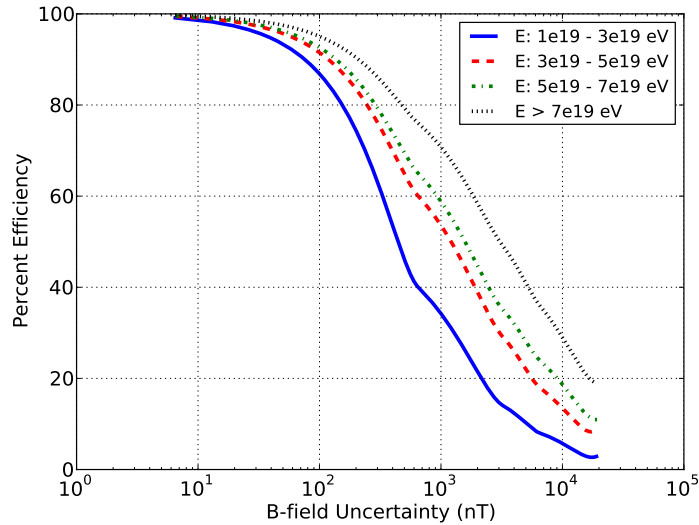}
\caption{Efficiency of real-time dedispersion trigger assuming deconvolution template parameter errors in the geomagnetic field parameter. For this test the TEC is assumed true which the amplitude and direction of the geomagnetic field value deviates from the true value. The current IGRF~\cite{igrf} model has geomagnetic field errors of 10~nT where the trigger efficiency is high.} 
\label{fig:geomag_efficiency}
\end{figure}

A similar analysis to determine the sensitivity of dedispersion to uncertainties in the geomagnetic field amplitude and directions is shown in Figure~\ref{fig:geomag_efficiency}.  Knowledge of the geomagnetic field needs to be within 400~nT for 50\% efficiency. The current IGRF models~\cite{igrf} have errors of 10~nT making it more than adequate for dispersion template modeling. The stability of the ionosphere is critical for proper functioning of SWORD and periods of high solar activity, which can cause large fluctuations in TEC and magnetic field values, should be avoided.

The duty cycle of the real-time dedispersion trigger is expected to be affected by periods of high TEC. Figure~\ref{fig:iono_distrib} shows the TEC variation with orbit. The TEC rises to very high levels for 10-25\% of the orbit time. The SWORD payload would continue to operate in these intervals but the trigger efficiency is expected to be affected.

\section{Outlook and Conclusions}
We have provided the scientific motivation, modeling, and expected performance for an orbiting UHECR detector using radio techniques that can provide the sensitivity required to open a new window to charged particle astronomy. Using the Earth's surface as a reflector achieves high exposure to energies $>10^{20}$~eV where cosmic rays are rigid enough to propagate from their sources undeflected. Full sky coverage will provide maps of cosmic ray events with several hundred events where only a handful have ever been detected.

Identification of a clear source of cosmic rays at extremely high energies, combined with observations of ultra-high energy neutrinos and gamma rays, would open a new understanding of cosmic accelerators. This data could solve the 50 year puzzle of the source and character of ultra-high energy cosmic ray accelerators. It is worth mentioning that sky maps of cosmic rays over the energy range observable by SWORD would provide information of strength and coherence length of extragalactic magnetic fields~\cite{stanev_1997}.  

For SWORD to succeed real-time dedispersion capabilities need to be developed. Although the computations themselves are not complicated, this study has found that they have to be performed at a high rate. We expect that prototyping the algorithms with field-programmable gate arrays and implementation in application-specific integrated circuits will provide the computational capabilities required within the power consumption constraints of the spacecraft.

\section{Acknowledgments}
We would like to thank all the participants of the mission concept study held at the Jet Propulsion Laboratory. These include Anthony Mannucci, Xiaoqing Pi, Joan Ervin, Dan Scharf, Kim Aaron, Bill Imbriale, Doug Lisman, Jeff Hilland, Tony Freeman, Matt Bennett, Melissa Vick, Asif Ahmed, Cece Guair, Steven Hu, Yoaz Bar-Sever, Charlie Dunn, Mark Thomson, Robert Navarro, and Larry D'Addario. This research was carried out at the Jet Propulsion Laboratory, California Institute of Technology, under a contract with the National Aeronautics and Space Administration. Copyright 2013. All rights reserved. 


\bibliographystyle{elsarticle-num}
\bibliography{<your-bib-database>}




\end{document}